\documentclass[12pt,titlepage]{utarticle-dan}

\usepackage{amsmath,amsthm,amsfonts,amscd,amssymb} 
\usepackage{eucal}
\usepackage{braket}
\usepackage{slashed}
\usepackage[hidelinks]{hyperref}
\usepackage{graphicx}
\usepackage{xcolor}
\usepackage{enumitem}
\usepackage{cite}

\usepackage{authblk}
\usepackage{tikz} 
\usetikzlibrary{shapes,arrows,positioning,automata,backgrounds,calc,er,patterns,decorations.markings,snakes}

\tikzset{->-/.style={decoration={
  markings,
  mark=at position #1 with {\arrow[scale=2]{>}}},postaction={decorate}}}
 
\tikzset{-<-/.style={decoration={
  markings,
  mark=at position #1 with {\arrow[scale=2]{<}}},postaction={decorate}}}

\tikzset{
photon/.style={decorate, draw=black,
    decoration={coil,aspect=0}}
 }

\tikzset{
gluon/.style={decorate, draw=black,
    decoration={coil,amplitude=4pt, segment length=5pt}}
 }

\newcommand{\be}{\begin{equation}}
\newcommand{\ee}{\end{equation}} 
\newcommand{\mb}{\mathbf}

\newcommand{\Oi}{\mathcal{O}}
\newcommand{\Li}{\mathcal{L}}

\newcommand{\ul}{\underline}

\newcommand\blfootnote[1]{%
  \begingroup
  \renewcommand\thefootnote{}\footnote{#1}%
  \addtocounter{footnote}{-1}%
  \endgroup
}

\title{Newton, entanglement, and the graviton}

    \author{Daniel Carney \\
	  Physics Division, Lawrence Berkeley National Laboratory \\
	  Berkeley, CA \\
	  {~}\\
	  \email{carney@lbl.gov}\\
       }

\Abstract{Many experiments have recently been proposed to test whether non-relativistic gravitational interactions can generate entanglement. In this note, I consider the extent to which these experiments can test if the graviton exists. Assuming unitarity and Lorentz invariance of the $S$-matrix, I demonstrate that this ``Newtonian entanglement'' requires the existence of massless bosons, universally coupled to mass, in the Hilbert space of low-energy scattering states. These bosons could be the usual spin-2 gravitons, but in principle there are other possibilities like spin-0 scalar gravitons. I suggest a concept for a more refined experiment to rule these out. The special role of $d=3+1$ spacetime dimensions and the possibility that unitarity is violated by gravity are highlighted.\blfootnote{Dedicated to the memory of Steve Weinberg (1933-2021).}}

\begin{document}

\maketitle

\tableofcontents

\section{Introduction}

Dyson and others have pointed out that detection of individual gravitons is likely to be impossible \cite{dysonorig,rothman2006can,dyson2013graviton}. Experimental proof of the existence of the graviton may thus require a more subtle methodology. 

\begin{figure}[t]

\begin{tabular}{cc}

\begin{tikzpicture}[scale=0.47]

\draw (0,3) -- (2,3);
\node at (-1.7,3) {$\ket{\psi_1}$};

\draw (0,1) -- (2,1);
\node at (-1.7,1) {$\ket{\psi_2}$};

\draw [black] (2,0.5) rectangle (6,3.5);
\node at (4,2) {$e^{-i V_N \Delta t}$};

\draw (6,1) -- (8,1);
\draw [black] (8,1.5) -- (8,.5) arc(-90:90:.5) -- cycle;

\draw (6,3) -- (8,3);
\draw [black] (8,3.5) -- (8,2.5) arc(-90:90:.5) -- cycle;

\node at (4,-1) {(a)};

\end{tikzpicture}

&

\begin{tikzpicture}[>=stealth',pos=.8,photon/.style={decorate,decoration={snake,post length=1mm}},scale=.96]

\draw [fill=lightgray] (-1.5,0) circle (.5);
\node at (-1.5,0) {$m_1$};
\draw [fill=lightgray] (1.5,0) circle (.5);
\node at (1.5,0) {$m_2$};

\draw [<->] (-.5,0) -- (.5,0);
\node at (0,.3) {$V_N$};


\draw (-4.5,.5) rectangle (-3.5,.1);
\node [style={scale=0.8}] at (-4,.3) {laser};
\draw [photon,->] (-3.5,.3) -- (-2,0);
\draw [photon,->] (-2,0) -- (-3.5,-.3);

\begin{scope}[shift={(-3.5,0)},rotate=180,scale=.3]
\draw (0,0.5) -- (1,1.5);
\draw (0,0.5) -- (1,0.5) -- (1,1.5) -- (0,1.5) -- (0,0.5);

\draw [black] (0,2.5) -- (1,2.5) arc(0:180:.5) -- cycle;
\draw (.5,1) -- (.5,2.5);

\draw [black] (2,1.5) -- (2,0.5) arc(-90:90:.5) -- cycle;
\draw (.5,1) -- (2,1);
\end{scope}


\begin{scope}[xscale=-1]

\draw (-4.5,.5) rectangle (-3.5,.1);
\node [style={scale=0.8}] at (-4,.3) {laser};
\draw [photon,->] (-3.5,.3) -- (-2,0);
\draw [photon,->] (-2,0) -- (-3.5,-.3);

\begin{scope}[shift={(-3.5,0)},rotate=180,scale=.3]
\draw (0,0.5) -- (1,1.5);
\draw (0,0.5) -- (1,0.5) -- (1,1.5) -- (0,1.5) -- (0,0.5);

\draw [black] (0,2.5) -- (1,2.5) arc(0:180:.5) -- cycle;
\draw (.5,1) -- (.5,2.5);

\draw [black] (2,1.5) -- (2,0.5) arc(-90:90:.5) -- cycle;
\draw (.5,1) -- (2,1);
\end{scope}

\end{scope}

\node at (0,-1.5) {(b)};

\end{tikzpicture}

\end{tabular}

\caption{(a) Circuit diagram of the simplest possible experimental implementation. Two masses $m_1,m_2$ are prepared in an initial product state, interact gravitationally for a time $\Delta t$, and are read out to check for entanglement. (b) Realization with free-falling masses. The Newton interaction would squeeze the relative position $x_-$ while preserving total momentum $p_+$, leading to a violation of the Duan inequality \cite{duan2000inseparability,giovannetti2003characterizing} for separable states $\braket{\Delta x_-^2 \Delta p_+^2} \geq \hbar^2$. This can be read out with local interferometers.}
\label{figure-cartoon}

\end{figure}
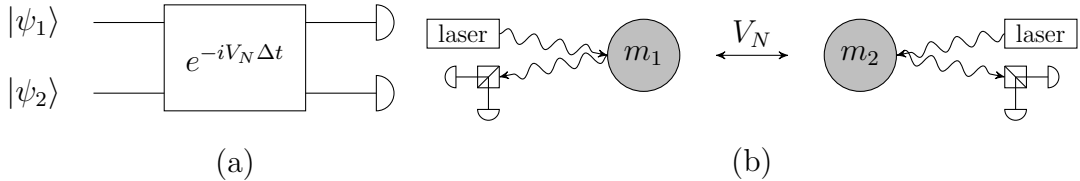

Bell's theorem provides a method to prove that a state of nature does not admit a local, classical description \cite{bell1964einstein,freedman1972experimental,aspect1981experimental}. There have been a plethora of experimental proposals to determine if gravitational interactions can generate such a non-classical, entangled state \cite{carney2019tabletop,cecile2011role,feynman1971lectures,page1981indirect,kafri2013noise,Bahrami:2015wma,anastopoulos2015probing,bose2017spin,marletto2017gravitationally,haine2021searching,Qvarfort:2018uag,Carlesso:2019cuh,howl2021non,Matsumura:2020law,Pedernales:2021dja,Carney:2021tsk,liu2021gravitational,Datta:2021ywm}. A minimal realization is depicted in Fig. \ref{figure-cartoon}.

These are non-relativistic tabletop experiments. If we assume that gravity generates a unitary channel on the objects, then observation of entanglement generation consistent with a $1/r$ law would tell us that the Hamiltonian must be
\be
\label{Hnon-rel}
H = H_1 + H_2 + V_{N}, \ \ \ V_{N} = - \frac{G_N m_1 m_2}{|\mb{x}_1 - \mb{x}_2 |}.
\ee
It is natural to ask \cite{Anastopoulos:2018drh,belenchia2018quantum,Galley:2020qsf,christodoulou2019possibility,marshman2020locality,Rydving:2021qua} what we learn about any gravitational degrees of freedom themselves, which do not appear in \eqref{Hnon-rel}. Of course, quantizing metric fluctuations into gravitons produces a perfectly good effective quantum field theory, and this model reproduces the Newton potential operator \cite{Donoghue:1994dn,Donoghue:1995cz,Burgess:2003jk}. The question these experiments probe is whether this is the correct model of nature.

In this paper, I study the implications if we further assume that \eqref{Hnon-rel} is the non-relativistic limit of some Lorentz-invariant model. A minimal $S$-matrix theory framework is sufficient to encode unitarity and Lorentz invariance without assuming that the fundamental degrees of freedom are quantum fields \cite{mandelstam1958determination,chew1960theory,Weinberg:1964ew,eden2002analytic}. These assumptions are enough to prove that one needs massless bosonic degrees of freedom in the Hilbert space of scattering states to be consistent with \eqref{Hnon-rel}. These bosons must have a universal coupling to mass with strength $\sqrt{G_N} m$. They can however have any integer spin, not only $s = 2$ like the usual graviton. Ruling out these other possibilities requires a more refined experiment, as discussed in section \ref{section-implications}.

Perhaps the more interesting possibility is that gravity does not generate a unitary interaction. For example, gravity could emerge from interactions mediated by unobserved microscopic degrees of freedom \cite{jacobson1995thermodynamics,Padmanabhan:2002xm,Verlinde:2010hp,Banks:2001px,Kafri:2014zsa,Jacobson:2015hqa,Banks:2020zcr}. Models of this type in which gravity emerges in a semi-classical fashion like $G_{\mu\nu} = 8 \pi G_N \braket{T_{\mu\nu}}$ would be ruled out, since there gravity cannot entangle objects \cite{carney2019tabletop}. However, it may be possible that gravity could form an open system in a different manner, in which it can produce entanglement observables consistent with \eqref{Hnon-rel}. Making a precise statement about gravitons in this context is a difficult problem left to future work.

Before moving on, we note some previous results in this direction \cite{belenchia2018quantum,Galley:2020qsf}. In particular, Belenchia et al. \cite{belenchia2018quantum} study a gedankanexperiment in which Newtonian entanglement enables superluminal signaling, and resolve the paradox by introducing quantized metric fluctuations. The arguments presented here are related, but precise enough to demonstrate an exhaustive list of possibilities: the \emph{only} way to resolve these types of paradoxes within a unitary and Lorentz-invariant model is to include radiative graviton, or very graviton-like, degrees of freedom.

\section{Unitary Newtonian gravity requires quantized radiation}
\label{section-main}

Our core assumption is unitarity. Specifically, we assume that the gravitational interaction between massive objects operates as a closed system. Within a unitary framework, a demonstration of entanglement between masses, with the right parametric dependences, means that we must have a two-body potential of the form
\be
\label{Hyukawa}
V_N = -\frac{G_N m^2 e^{-\mu |\mb{x}_1 - \mb{x}_2|}}{|\mb{x}_1 - \mb{x}_2|}.
\ee
Here $\mu$ is a small parameter with dimensions of inverse length which is useful as a regulator; we will take the $\mu \to 0$ limit to recover the Newton potential.\footnote{The experiments are done within some finite length scale, typically a laboratory scale $10^{-6}~{\rm m} \lesssim \ell \lesssim 1~{\rm m}$. Thus beyond its regulatory benefits, this kind of potential would be a perfectly viable explanation for the observed entanglement as long as we assume a sufficiently small value $\mu \ll \ell^{-1}$. For extensive discussion on the $S$-matrix in the $\mu \to 0$ limit, see for example \cite{dollard1964asymptotic,weinberg1965infrared,kulish1970asymptotic,carney2017infrared,Caron-Huot:2021rmr}.} The real input from the experiments is that this is a two-body operator in the full sense of the term: the $\mb{x}_i$ are local operators on a bipartite Hilbert space. This is different from, for example, models where gravity acts ``semiclassically'' through expectation values (see appendix \ref{appendix-classical}).

We will consider scattering processes with the Hamiltonian \eqref{Hyukawa}. Using the non-relativistic matrix elements as input to Lorentz-invariant extensions of this model, we will find violations of unitarity in a variety of simple scattering processes. Moreover, the precise form of this unitarity violation is enough to conclude that the only solution is to add radiative states of bosons with mass $\mu$ which couple to massive matter with strength $\sqrt{G_N} m$.

Unitarity and Lorentz invariance can be defined in a manner which is independent of the way we realize the interactions, i.e., does not depend on the use of field theory. Consider scattering processes described by $S$-matrix elements $S_{\alpha \to \beta} = \braket{\beta | S | \alpha}$. The $S$-matrix elements are the transition amplitudes for an initial state $\ket{\alpha}$ prepared in the asymptotic past $t \to -\infty$ to evolve to a definite state $\ket{\beta}$ in the asymptotic future $t \to +\infty$. To begin, we will only need the minimal postulates:
\begin{enumerate}[label=(\Alph*)]

\item \label{assumption-unitarity} {\bf Unitarity}. $S$ is a unitary operator, $S^{\dagger}S = 1$.

\item \label{assumption-lorentz} {\bf Lorentz invariance}. For $\Lambda$ an element of the Lorentz group, the scattering states transform in a unitary representation $U(\Lambda)$. Furthermore, the $S$-matrix is invariant: $U(\Lambda) S U^{\dagger}(\Lambda) = S$. 

\end{enumerate}
Our implementation of \eqref{Hyukawa} to compute $S$-matrix elements non-relativistically will be based on a Schr\"{o}dinger equation with past boundary conditions, and thus automatically assumes the usual non-relativistic notion of causality. 

Gravity is a weak interaction, so that we can expand $S = 1 + i T$, where the $1$ reflects the possibility of no scattering occurring. Assumption (B) means in particular that a spinless particle of momentum $\mb{p}$ transforms like $U(\Lambda) \ket{\mb{p}} = \sqrt{(\Lambda p)^0/p^0} \ket{\Lambda \mb{p}}$ under a boost.\footnote{This is our \emph{definition} of a single-particle state; it could be a field quanta, or a closed string mode, or whatever else, as long as it transforms correctly. Following \cite{weinberg1995quantum}, we define single-particle states to satisfy the non-relativistic normalization $\braket{\mb{p}' | \mb{p}} = \delta^3(\mb{p}'-\mb{p})$, and use $(-,+,+,+)$ signature.} Multiple-particle states $\ket{\alpha} = \ket{\mb{p}_1 \mb{p}_2 \cdots}$ are described by lists of such momenta (as well as spins and any other internal quantum numbers). Furthermore, spacetime translation invariance implies that total four-momentum is conserved in every process. Thus we will define the usual ``Feynman amplitudes'' $M$ by
\be
\label{feynman}
S_{\alpha \to \beta} = \delta_{\alpha\beta} - i (2 \pi)^4 \delta^4(p_{\alpha} - p_{\beta}) B_{\alpha} B_{\beta} M_{\alpha \to \beta},
\ee
where $p_{\alpha}, p_{\beta}$ are the total incoming and outgoing four-momenta, respectively. We will only deal with spinless massive objects, and so have factored out the Lorentz-transformation factors $B_{\alpha} = \prod_{i \in \alpha} [2(2\pi)^3 E_{i}]^{-1/2}$, with $E_i = p_i^0$ the energy of the $i$th particle. Defined this way, $M_{\alpha \to \beta}$ should be invariant under the Lorentz group.

The main workhorse in what follows will be the unitarity condition on the $S$ matrix. We have $S^{\dagger} S = 1$, which implies $i (T - T^{\dagger}) = T^{\dagger} T$. Inserting a complete set of final states $\ket{X}$ and comparing to \eqref{feynman}, we have
\be
\label{optical}
i \left( M_{\alpha \to \beta} - M^*_{\beta \to \alpha} \right) = (2\pi)^4 \sum_{X} B_X^2 M_{\alpha \to X} M^*_{\beta \to X} \delta^4(p_{\alpha} - p_X).
\ee
This is known as the optical theorem. In the special case of forward scattering $\alpha = \beta$, this reduces to the usual optical theorem ${\rm Im}~f(0) = \sigma_{\rm total}$. This is a reflection of the fact that the scattered wavefunction must have a specific interference pattern with the unscattered wavefunction. For a model to be unitary, equation \eqref{optical} must be satisfied for all initial and final states $\ket{\alpha}$, $\ket{\beta}$.

\subsection{Lorentzian bootstrap strategy}
\label{section-bootstrap}

\begin{figure}[t]

\centering

\begin{tikzpicture}[scale=.9,every node/.style={scale=0.9}]

\draw [->] (-2.2,-1.8) -- (-2.2,1.8);
\node at (-2.5,0) {\rotatebox{90}{time}};

\draw [pattern=north west lines] (0,0) circle (.5);
\draw (0,.5) -- (0,1.5);
\draw (0,-.5) -- (0,-1.5);
\draw (.35,.35) -- (1,1.5);
\draw (-.35,.35) -- (-1,1.5);
\draw (-.35,-.35) -- (-1,-1.5);
\draw (.35,-.35) -- (1,-1.5);

\draw [decoration={brace},decorate] (-1,1.7) -- (1,1.7);
\node at (0,2.2) {$\ket{\beta}$};

\draw [decoration={brace},decorate] (1,-1.7) -- (-1,-1.7);
\node at (0,-2.2) {$\ket{\alpha} = \ket{\mb{p}_1 \mb{p}_2 \cdots }$};

\node at (0,-3) {(a)};

\begin{scope}[xshift=150]

\draw (-2,-1) -- (-1,0);
\draw (-2,1) -- (-1,0);
\draw [fill=lightgray] (0,0) ellipse [x radius=1, y radius=.1];
\draw (1,0) -- (2,1);
\draw (1,0) -- (2,-1);

\node at (0,.4) {$V_N(t)$};

\node at (-2,1.3) {$\mb{p}'_1$};
\node at (2,1.3) {$\mb{p}'_2$};
\node at (-2,-1.3) {$\mb{p}_1$};
\node at (2,-1.3) {$\mb{p}_2$};

\node at (0,-3) {(b)};

\begin{scope}[xshift=180];

\draw (-2,-1) -- (-1,0);
\draw (-2,1) -- (-1,0);
\draw [dashed] (-1,0) -- (1,0) ;
\draw (1,0) -- (2,1);
\draw (1,0) -- (2,-1);

\node at (-2,1.3) {$\mb{p}'_1$};
\node at (2,1.3) {$\mb{p}'_2$};
\node at (-2,-1.3) {$\mb{p}_1$};
\node at (2,-1.3) {$\mb{p}_2$};

\node at (0,-3) {(c)};

\end{scope}

\end{scope}

\end{tikzpicture}

\caption{(a) Depiction of an $S$-matrix element for the process $\ket{\alpha} \to \ket{\beta}$. The blob represents a sum over all possible intermediate processes. (b) Diagram describing the gravitational, non-relativistic scattering amplitude at lowest order in perturbation theory. The interaction is given by the instantaneous potential matrix element $ \sim [ (\mb{p}'_1 - \mb{p}_1)^2 + \mu^2]^{-1}$. (c) Lorentzian bootstrap version of the same amplitude. The dashed line represents a factor $[ (p_1'-p_1)^2 + \mu^2]^{-1}$, which at this stage in the argument has no interpretation in terms of an intermediate particle.}
\label{figure-treelevel}
\end{figure}
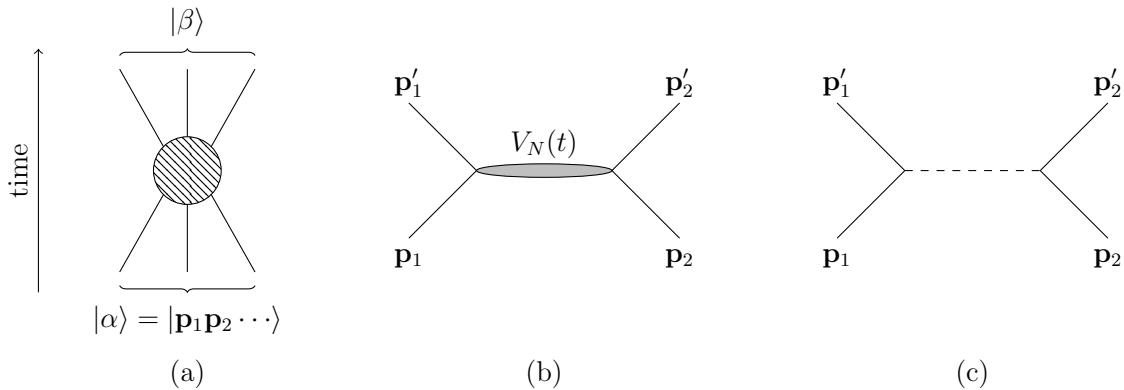

The strategy we will follow is to begin with a non-relativistic expression for the scattering amplitude, then bootstrap the answer into something Lorentz-invariant.\footnote{The ``bootstrap'' terminology is inspired from the modern $S$-matrix bootstrap program, where one derives constraints on low-energy models by demanding that they are low-energy limits of some high-energy model with certain restrictions \cite{Adams:2006sv}. Usually this is used to relate a pair of relativistic field theories, but for a few examples with non-relativistic models, see \cite{Pajer:2020wnj,Grall:2021xxm,Mojahed:2021sxy}.} Scattering theory is the study of solutions to the Schr\"{o}dinger equation combined with boundary conditions as $t \to -\infty$. Our starting point will thus be the time-dependent perturbation series solution for this system. Although the Born series is perhaps more familiar, consider instead the Dyson series \cite{weinberg2015lectures}
\be
\label{dyson}
S_{\alpha \to \beta} = \braket{\beta | S | \alpha} = \sum_{n=0}^{\infty} \frac{(-i)^n}{n!} \int_{-\infty}^{\infty} dt_1 \cdots dt_n \braket{\beta | \mathcal{T} V_I(t_1) \cdots V_I(t_n) | \alpha}.
\ee
Here $\mathcal{T}$ is the time-ordering operator, and the potential is written in the interaction picture. To illustrate the Lorentzian bootstrap idea, consider $2\to 2$ scattering, so $\ket{\alpha} = \ket{ \mb{p}_1 \mb{p}_2}$ and $\ket{\beta} = \ket{ \mb{p}'_1 \mb{p}'_2}$. We will write the interaction picture potential operator \eqref{Hyukawa} explicitly in two-body notation, in the lab frame:
\begin{align}
\begin{split}
\label{VN-int}
V_N(t) = & \ -\frac{1}{(2\pi)^{3}} \int d^3\mb{k}_1 d^3\mb{k}_2 d^3\mb{k}'_1 d^3\mb{k}'_2 \delta^3(\mb{k}_1 + \mb{k}_2 - \mb{k}'_1 - \mb{k}'_2) \\
& \times e^{-i (E_{\mb{k}'_1
}+E_{\mb{k}'_2} - E_{\mb{k}_1} - E_{\mb{k}_2})t} 
V_{\mb{k}'_1 \mb{k}'_2,\mb{k}_1 \mb{k}_2} \ket{\mb{k}'_1 \mb{k}'_2} \bra{\mb{k}_1 \mb{k}_2},
\end{split}
\end{align}
where the Schr\"{o}dinger-picture matrix elements are
\be
\label{Velements}
V_{\mb{k}'_1 \mb{k}'_2,\mb{k}_1 \mb{k}_2} = \frac{4\pi G_N m^2}{ \mb{q}^2 + \mu^2}, \ \ \ \mb{q} =\frac{\mb{k}_1' - \mb{k}_2'}{2} -  \frac{\mb{k}_1 - \mb{k}_2}{2},
\ee
and $\mb{q}$ is the change in the relative momentum. The zeroth order term $n=0$ in \eqref{dyson} is just the identity operator in the expansion $S = 1 + i T$. The first-order term gives the same result as the first Born approximation:
\be
\label{S1non-rel}
S^{(1)}_{\mb{p}_1 \mb{p}_2 \to \mb{p}'_1 \mb{p}'_2} = \frac{2\pi i}{(2\pi)^3} \delta(E_{\mb{p}_1
}+E_{\mb{p}_2} - E_{\mb{p}'_1} - E_{\mb{p}'_2}) \delta^3(\mb{p}_1 + \mb{p}_2 - \mb{p}'_1 - \mb{p}'_2) \frac{4\pi G_N m^2}{\Delta \mb{p}^2 + \mu^2},
\ee
with $\Delta \mb{p} = \mb{p}_1' - \mb{p}_1$ the momentum transfer. 

Now we impose the requirement that \eqref{S1non-rel} is the non-relativistic limit of some Lorentz-invariant model. Thus we need to write this as Lorentz-invariant function of the external momenta $\mb{p}_1, \mb{p}_2, \mb{p}'_1, \mb{p}'_2$. Specifically, we want a Feynman amplitude $M$ with the property that $S^{(1)} = \lim_{p^0 \to m} -i(2\pi)^4 M \delta^4(\sum p)$. There are 10 Lorentz generators and 12 independent variables in these four momenta, leaving only two free variables. A convenient choice for these are the invariant momentum transfer $t = -(p_1' - p_1)^2$ and total incoming mass-squared $s = -(p_1 + p_2)^2$. Clearly the only Lorentz-invariant option is to replace the momentum transfer $\Delta \mb{p}^2$ with $t$, and combine the two delta functions into a factor $\delta^4(p_1 + p_2 - p_1' - p_2')$. Comparing to \eqref{feynman}, we have the bootstrapped amplitude
\be
\label{M1rel}
M_{\mb{p}_1 \mb{p}_2 \to \mb{p}'_1 \mb{p}'_2} = -\frac{16\pi G_N m^4}{-t + \mu^2}.
\ee
We had to rescale the coupling $G_N m^2 \to G_N m^4$ in order to cancel the relativistic wavefunction normalizations $\sim 1/\sqrt{p^0} \to 1/\sqrt{m}$ in \eqref{feynman}. The two expressions for the amplitude are depicted in Fig. \ref{figure-treelevel}. We have assumed that the two massive particles are distinguishable so that we can ignore the exchange channel $\mb{p}'_1 \leftrightarrow \mb{p}'_2$, and will continue to do this in what follows.

Unitarity is trivially satisfied in \eqref{M1rel}. The amplitude is manifestly real, so the left-hand side of \eqref{optical} vanishes. Similarly, there is no amplitude at order $\sqrt{G_N}$, so to order $G_N$ the right hand side also vanishes. Notice that for physical momenta $p_1^2 = p_2^2 = p_1'^2 = p_2'^2 = -m^2$, we have $t \leq 0$, so the denominator is always non-zero.

To see how the unitarity condition can become non-trivial, we have to go to higher order in perturbation theory. We next show an example of how this works. We consider a $3 \to 3$ process which includes two non-gravitational interactions with coupling strength $\lambda$. This is modeled directly after the kind of Alice and Bob experiments of \cite{belenchia2018quantum}. In this example, the unitarity violation shows up at order $G_N \lambda^2$. To emphasize the generality of the core idea, we also show a purely gravitational example in appendix \ref{appendix-particleantiparticle}. There we consider a $2 \to 2$ process with an incoming particle-antiparticle pair, in which unitarity violations arise at order $G_N^2$.

\subsection{Tree-level unitarity}
\label{section-tree}

\begin{figure}[t]
\centering

\begin{tikzpicture}[scale=.9,every node/.style={scale=0.9}]

\draw (-2,-1) -- (-1,0);
\draw (-2,1) -- (-1,0);
\draw [fill=lightgray] (0,0) ellipse [x radius=1, y radius=.1];
\draw (1,0) -- (2,1);
\draw (1,0) -- (2,-1);
\draw [photon] (-1.5,-2) -- (-1.5,-.5);
\draw [photon] (1.5,2) -- (1.5,.5);

\node at (0,.4) {$V_N(t)$};

\node at (-2,1.3) {$\mb{p}'_1$};
\node at (2,1.3) {$\mb{p}'_2$};
\node at (-2,-1.3) {$\mb{p}_1$};
\node at (2,-1.3) {$\mb{p}_2$};
\node at (-1.2,-2) {$\mb{k}$};
\node at (1.2,2) {$\mb{k}'$};

\node at (0,-3) {(a)};

\begin{scope}[xshift=210]

\draw (-2,-1) -- (-1,0);
\draw (-2,1) -- (-1,0);
\draw [dashed] (-1,0) -- (1,0); 
\draw (1,0) -- (2,1);
\draw (1,0) -- (2,-1);
\draw [photon] (-1.5,-2) -- (-1.5,-.5);
\draw [photon] (1.5,2) -- (1.5,.5);

\node at (-2,1.3) {$\mb{p}'_1$};
\node at (2,1.3) {$\mb{p}'_2$};
\node at (-2,-1.3) {$\mb{p}_1$};
\node at (2,-1.3) {$\mb{p}_2$};
\node at (-1.2,-2) {$\mb{k}$};
\node at (1.2,2) {$\mb{k}'$};

\node at (0,.3) {$\tilde{k}$};
\node at (-2.4,.2) {$p_1+k$};
\draw [->] (-2,0) -- (-1.3,-.2);
\node at (2.7,0) {$p_2'+k'$};
\draw [->] (2,0) -- (1.3,.2);

\node at (0,-3) {(b)};

\end{scope}

\end{tikzpicture}

\caption{Scattering amplitude for the $3 \to 3$ process with external photons, in the same notation as Fig. \ref{figure-treelevel}. (a) is the non-relativistic amplitude and (b) is its relativistic extension.}
\label{figure-6pt}

\end{figure}
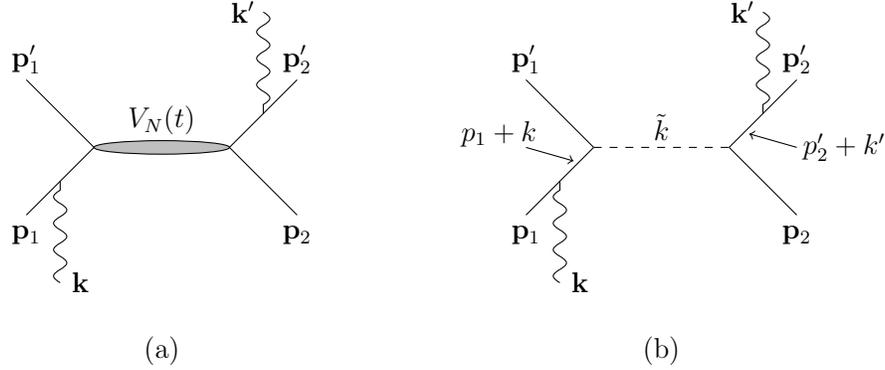

Consider a process where two experimentalists Alice and Bob each have a massive particle. At some early time Bob interacts with his particle, say by hitting it with a photon. The two particles proceed to scatter via the Newton interaction. Long after this scattering event, Alice then performs a measurement of her particle, say again by hitting it with a photon. Let us model the photon interactions with an interaction strength $\lambda$ (proportional to the charge of Alice and Bob's particles). Following the logic of the Lorentzian bootstrap given above, one obtains an amplitude
\be
\label{M-tree}
M_{\mb{k} \mb{p}_1 \mb{p}_2 \to \mb{k}' \mb{p}_1' \mb{p}_2'} = \left( \frac{\lambda}{(p_1 + k)^2 + m^2 - i \epsilon}\right) \left(  \frac{G_N m^4}{\tilde{k}^2 + \mu^2 - i \epsilon} \right) \left( \frac{\lambda}{(p_2' + k')^2 + m^2 - i \epsilon} \right).
\ee
This is depicted diagrammatically in Fig. \ref{figure-6pt}. The photon momenta are $\mb{k},\mb{k}'$ respectively. The four-momentum transfer between Alice and Bob's massive particles is now $\tilde{k} = p_1' - (p_1 + k)$. See appendix \ref{appendix-details} for some details of this calculation.

Crucially, it is now possible that $\tilde{k}^2 = -\mu^2$. This happens when the photon momentum $|\mb{k}| \gtrsim \mu$. This is why we have inserted the $i\epsilon$ factors in the denominators: the poles are now accessible to the experimentalist. When we use the $S$-matrix to describe an actual scattering experiment, it needs to be integrated against some momentum-space wavepackets describing the incoming and outgoing particles. To make these integrals well-defined, we have to say what happens on the poles. The specific prescription used here (the ``Feynman prescription'') is the only possibility consistent with our unitarity and Lorentz-invariance assumptions, as explained in appendix \ref{appendix-details}.

At points in phase space where $\tilde{k}^2 = -\mu^2$, the pole in the middle term has an imaginary residue, because 
\be
\lim_{\epsilon \to 0} \frac{1}{x - i \epsilon} = \frac{1}{x} - i \pi \delta(x)
\ee
in the sense of distributions (for $x$ real). Therefore, unitarity is no longer trivial: the left-hand side of the optical theorem \eqref{optical} is now non-zero. Specifically, it has a pole singularity at $\tilde{k}^2 = -\mu^2$ with residue of order $\lambda^2G_N $. Therefore, if unitarity were to hold, we would need to have an amplitude $M \sim \Oi(\lambda \sqrt{G_N})$ to use in the right-hand side of \eqref{optical}. But if the full set of scattering states is only massive particles and photons, it is easy to see that there simply is no such amplitude!\footnote{One could distribute the coupling factors differently between the two amplitudes on the right-hand side of \eqref{optical}. In fact there are some disconnected diagrams like this with the correct scalings of the couplings, but they have the wrong detailed momentum dependence.} Therefore the optical theorem fails and we have a violation of unitarity.

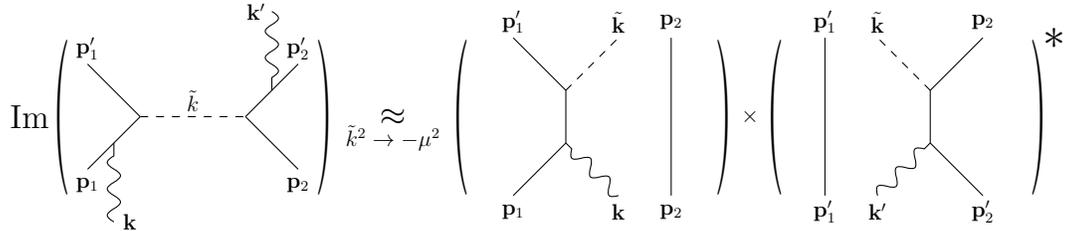
\begin{figure}[t]

\centering
\begin{tikzpicture}[scale=.7,every node/.style={scale=0.7}]

\node [scale=1.5] at (-3.1,0) {Im};
\node [yscale=7,xscale=2] at (-2.5,0) {(};
\node [yscale=7,xscale=2] at (2.5,0) {)};

\draw (-2,-1) -- (-1,0);
\draw (-2,1) -- (-1,0);
\draw [dashed] (-1,0) -- (1,0); 
\draw (1,0) -- (2,1);
\draw (1,0) -- (2,-1);
\draw [photon] (-1.5,-2) -- (-1.5,-.5);
\draw [photon] (1.5,2) -- (1.5,.5);

\node at (-2,1.3) {$\mb{p}'_1$};
\node at (2,1.3) {$\mb{p}'_2$};
\node at (-2,-1.3) {$\mb{p}_1$};
\node at (2,-1.3) {$\mb{p}_2$};
\node at (-1.2,-2) {$\mb{k}$};
\node at (1.2,2) {$\mb{k}'$};
\node at (0,.3) {$\tilde{k}$};

\node[scale=1.5] at (3.8,0) {$\approx$};
\node[scale=1] at (3.8,-.4) {$\tilde{k}^2 \to -\mu^2$};

\begin{scope}[xshift=230]

\node [yscale=7,xscale=2] at (-3,0) {(};
\node [yscale=7,xscale=2] at (2,0) {)};

\draw (-2,-1.5) -- (-1,-.5);
\draw [photon] (0,-1.5) -- (-1,-.5);
\draw (-1,-.5) -- (-1,.5);
\draw (-1,.5) -- (-2,1.5);
\draw [dashed] (-1,.5) -- (0,1.5);

\draw (1,-1.5) -- (1,1.5);

\node at (-2,1.8) {$\mb{p}'_1$};
\node at (1,1.8) {$\mb{p}_2$};
\node at (-2,-1.8) {$\mb{p}_1$};
\node at (1,-1.8) {$\mb{p}_2$};
\node at (0,-1.8) {$\mb{k}$};
\node at (0,1.8) {$\tilde{\mb{k}}$};

\node at (2.5,0) {$\times$};

\end{scope}

\begin{scope}[xshift=400]

\node [yscale=7,xscale=2] at (-3,0) {(};
\node [yscale=7,xscale=2] at (2,0) {)};
\node [scale=2] at (2.3,1.5) {$*$};

\begin{scope}[xscale=-1,xshift=30]

\draw (-2,-1.5) -- (-1,-.5);
\draw [photon] (0,-1.5) -- (-1,-.5);
\draw (-1,-.5) -- (-1,.5);
\draw (-1,.5) -- (-2,1.5);
\draw [dashed] (-1,.5) -- (0,1.5);

\draw (1,-1.5) -- (1,1.5);

\node at (-2,1.8) {$\mb{p}_2$};
\node at (1,1.8) {$\mb{p}_1'$};
\node at (-2,-1.8) {$\mb{p}_2'$};
\node at (1,-1.8) {$\mb{p}_1'$};
\node at (0,-1.8) {$\mb{k}'$};
\node at (0,1.8) {$\tilde{\mb{k}}$};
\end{scope}

\end{scope}

\end{tikzpicture}

\caption{Unitarity in the tree-level, six-point amplitude. The pole at $\tilde{k}^2 = -\mu^2$ has an imaginary residue. The form of this residue is exactly equal to the product of a pair of amplitudes where the ``graviton'' line (dashed) is emitted into the final state. The disconnected lines represent trivial propagators.}
\label{figure-treeunitarity}
\end{figure}

What happened? The unitarity violation means that an incoming 3-body state $\ket{\psi} = \ket{\mb{k} \mb{p}_1 \mb{p}_2}$ will evolve to some state $\ket{\psi'} = S \ket{\psi}$ with a deficit in its norm $|\braket{\psi ' | \psi '}| < 1$. Inspecting the form of \eqref{optical}, and remembering that the dynamics are fixed by assumption of \eqref{Hyukawa}, we see that there is only one possible resolution: modify the sum over final states $\ket{X}$. In more detail, notice that near the pole $\tilde{k}^2 = -\mu^2$, the imaginary part of the amplitude behaves like
\begin{equation}
{\rm Im}~M_{\mb{k} \mb{p}_1 \mb{p}_2 \to \mb{k}' \mb{p}_1' \mb{p}_2'} \to G_N m^4 \left( \frac{\lambda}{(p_1 + k)^2 + m^2}\right) \delta(\tilde{k}^2 + \mu^2) \left( \frac{\lambda}{(p_2' + k')^2 + m^2} \right).
\end{equation}
But this is essentially just the product of two amplitudes with final state $\ket{X} = \ket{\tilde{\mb{k}} \mb{p}_1' \mb{p}_2}$, where the $\tilde{\mb{k}}$ represents a radiated particle of mass $\mu$ coupled with strength $\sqrt{G_N} m$. See Fig. \ref{figure-treeunitarity}. To be precise, if we include such final states, the optical theorem \eqref{optical} will be satisfied
\begin{equation}
\label{opticalfinal}
i( M_{\mb{k} \mb{p}_1 \mb{p}_2 \to \mb{k}' \mb{p}_1' \mb{p}_2'} - M^*_{\mb{k}' \mb{p}_1' \mb{p}_2' \to \mb{k} \mb{p}_1 \mb{p}_2}) \sim M_{\mb{k} \mb{p}_1 \mb{p}_2 \to \tilde{k} \mb{p}_1' \mb{p}_2} M^*_{\mb{k}' \mb{p}_1' \mb{p}_2' \to \tilde{k} \mb{p}_1' \mb{p}_2},
\end{equation}
where the sum on final states $\ket{X}$ has collapsed into a single outgoing state which includes this gravitationally-coupled radiation. (The $\sim$ represents some awkward factors arising from the disconnected propagators; see appendix \ref{appendix-details} for the detailed equality). Thus we conclude that we need to include such states in the Hilbert space of scattering states. In the limit $\mu \to 0$ this radiated particle is essentially a graviton in terms of its masslessness and $\sqrt{G_N}$ coupling to matter. It must be a boson, otherwise the diagrams on the right-hand side would violate angular momentum conservation. However, nothing in this argument is sensitive to which integer spin this boson carries.

The fact that we get a unitarity violation precisely when $\tilde{k}^2 = -\mu^2$ has a simple physical interpretation. In the language of field theory, $\tilde{k}^2 = -\mu^2$ occurs when the momentum transfer $\tilde{k} = p_1' - (p_1 + k)$ is tuned so that the ``virtual graviton'' mediating the interaction satisfies its relativistic dispersion relation, i.e., ``goes on shell''. This is why we need the external photon to see the effect: if $k = 0$ then $\tilde{k}^2 \geq 0$, but including the photon allows us to reach the pole at $\tilde{k}^2 = -\mu^2$.

This example bears some important similarities to the Alice and Bob gedankenexperiment of \cite{belenchia2018quantum}. There, causality (or rather faster-than-light signaling) was used as a primary diagnosis of the issues arising if one neglects to include final-state graviton radiation. Here we have instead focused on a scattering calculation, in which boundary conditions are imposed on both the past and future, which obscures the causal properties of the process. We found a unitarity violation instead of a causality violation. This reflects the fact that unitarity and causality are intimately linked in a relativistic model. Very similar considerations have long been discussed in the context of a classic paradox of Fermi \cite{fermi1932quantum}, who incorrectly argued that perturbation theory predicted superluminal communications. As is now well-known, the solution is that Fermi forgot to include final-state radiation (see \cite{dickinson2016probabilities} for a review).

It is interesting to note that the arguments given here for the necessity of a graviton-like particle depend strongly on the dimensionality of space-time. Consider general relativity in $d=2+1$ dimensions, defined as usual by the action
\be
S = \int d^3x \sqrt{-g} \left[ \frac{R}{16 \pi G_N} + \mathcal{L}_{\rm matter}\right].
\ee
This model is ``topological'' in the sense that is has no propagating gravitational waves \cite{deser1984three}. However, particles can pick up braiding phases while scattering \cite{witten19882+,carlip1989exact}, and thus can become entangled. The discrepancy with the argument above is that the Newton potential, or rather its logarithmic cousin in two spatial dimensions, is not the non-relativistic limit $d=2+1$ Einstein gravity \cite{deser1984three}. This exemplifies the fact that entanglement generation alone does not require a mediator---the key is the local form of the non-relativistic potential.

\section{Implications and interpretation}

\label{section-implications}

The bottom-up argument given above says that Newtonian entanglement can only be explained within a unitary, Lorentz-invariant model if that model includes radiative graviton-like degrees of freedom. Here I emphasize the fact that this does not uniquely pick out the spin-2 graviton by constructing some top-down counterexamples. A refined, non-Newtonian experiment is then outlined which could distinguish the spin of the gravitational mediator.

\subsection{Models which can explain the observation of entanglement}

We can certainly show that the graviton reproduces the necessary entangling operation on the masses. To see this, assume the metric is perturbatively expanded $g_{\mu\nu} = \eta_{\mu\nu} + \sqrt{32 \pi G_N} h_{\mu\nu}$ where $\eta_{\mu\nu}$ is flat spacetime. We have scaled out a factor of the Planck mass $m_{\rm pl} \sim 1/\sqrt{G_N}$ to give $h$ dimensions of mass. The graviton $h_{\mu\nu}$ couples to matter in the usual way
\be
\Li_{\rm int} = \sqrt{8\pi G_N} h^{\mu\nu} T_{\mu \nu} + \Oi(h^2)
\ee
with $T$ the stress-energy tensor. We will not need the terms quadratic and higher order in the gravitons. The Feynman rules for calculating amplitudes in this model are given in appendix \ref{appendix-details}. One finds the lowest-order $S$-matrix element \cite{Donoghue:1994dn,Donoghue:1995cz,Burgess:2003jk}
\be
\label{Smatrix-graviton}
M_{\mb{p}_1' \mb{p}_2', \mb{p}_1 \mb{p}_2} = 
\vcenter{\hbox{\begin{tikzpicture}
\draw (-1,-1) -- (0,0);
\node at (-1,-1.3) {$\mb{p}_1$};
\draw (-1,1) -- (0,0);
\node at (-1,1.3) {$\mb{p}_1'$};
\draw [gluon] (0,0) -- (2,0);
\node at (1,-.4) {$p_1'-p_1$};
\draw (2,0) -- (3,1);
\node at (3,1.3) {$\mb{p}_2'$};
\draw (2,0) -- (3,-1);
\node at (3,-1.3) {$\mb{p}_2$};
\end{tikzpicture}}} 
= 4\pi G_N \frac{N_2}{(p_1'-p_1)^2 - i \epsilon}
\ee
where the numerator
\begin{equation}
N_2 = 4 \left[ (p_1 \cdot p_2') (p_1' \cdot p_2) - m^2 (p_1 \cdot p_1') - m^2 (p_2 \cdot p_2') - 2 m^4 \right] \to 4 m^4
\end{equation}
in the non-relativistic limit $p^0 \to m, \mb{p} \to 0$ (so $p^2 \to -m^2$). This recovers the Newtonian result \eqref{M1rel}, in the $\mu \to 0$ limit. In other words, virtual graviton exchange leads to the Newtonian potential operator.

It is tempting to conclude from the above that observing Newtonian entanglement experimentally would imply that the graviton exists. But this is not correct. Nothing in any of the proposed experiments to date can detect the spin of the boson. 

To make this clear, let us study a simple model of scalar gravity \cite{Sundrum:2003yt}. Variants on this model go back to the days before general relativity \cite{nordstrom1913theorie}. Consider a scalar field $\phi$ coupled to matter through the trace of the stress tensor:
\be
\Li_{\rm int} = \sqrt{8\pi G_N} \phi T_{\mu}^{\mu}.
\ee
This coupling is Lorentz invariant and has the same mass dimensions as the usual graviton coupling. One way to obtain this would be to write the usual metric interaction but constrain the metric to be of the form $g_{\mu\nu} = \phi \eta_{\mu\nu}$ (``conformally flat''). This model is equivalent to Einstein gravity in the non-relativistic limit. To see this, consider the same $2 \to 2$ scattering matrix element:
\be
M_{\mb{p}_1' \mb{p}_2', \mb{p}_1 \mb{p}_2} = 
\vcenter{\hbox{\begin{tikzpicture}
\draw (-1,-1) -- (0,0);
\node at (-1,-1.3) {$p_1$};
\draw (-1,1) -- (0,0);
\node at (-1,1.3) {$p_1'$};
\draw [dashed] (0,0) -- (2,0);
\node at (1,-.4) {$p_1'-p_1$};
\draw (2,0) -- (3,1);
\node at (3,1.3) {$p_2'$};
\draw (2,0) -- (3,-1);
\node at (3,-1.3) {$p_2$};
\end{tikzpicture}}}  = 4\pi G_N \frac{N_0}{(p_1'-p_1)^2 - i \epsilon}.
\ee
The difference is the numerator,
\be
N_0 = 4 (p_1 \cdot p_1' + 2m^2) (p_2 \cdot p_2' + 2m^2) \to 4 m^4
\ee
which is in exact agreement with the spin-2 model in the non-relativistic limit. In other words, this spin-0 model produces the exact same effective Newtonian potential as the spin-2 model.

While this spin-0 mediator is a simple example, it is not the only possibility. In fact, one could construct a model like this using any integer spin. Even fermions are possible, in some sense. The force mediator could be condensed fermion pairs (like supercurrent fluctuations in a BCS superconductor), if one could think of a palatable way to violate the assumptions of the Weinberg-Witten theorem \cite{weinberg1980limits,Orland:2021wnp}.

\subsection{Measuring the mediator spin}
\label{section-spin}

This brings up the question: what would be needed to further pin down the spin? Of course, one could try to appeal to various classical observations, like the tensorial nature of gravitational waves detected by LIGO. It may be hard to imagine that the world has spin-2 classical radiation at astrophysical wavelengths but spin-0 radiation at tabletop scales. However, precisely this kind of situation could arise in a modified gravity scenario \cite{hoyle2001submillimeter,khoury2004chameleon} or a variety of dark matter models \cite{Hui:2016ltb}. More to the point, we are discussing experimental tests of quantum gravity, so one should be careful. 

\begin{figure}[t]
\centering

\begin{tikzpicture}

\draw [pattern=north west lines] (-1,0) circle (.4);
\draw [pattern=north west lines] (1,0) circle (.4);
\node at (-1,-1) {$\ket{L}$};
\node at (1,-1) {$\ket{R}$};

\draw [->-=.3] (3,2) -- (3,0);
\draw [->-=.7,bend left=10] (3,0) to (2.5,-2);
\draw [->-=.7,dashed,bend left=10] (3,0) to (2,-2);

\node at (3.4,1.5) {$\ket{0}$};
\node at (1.7,-2) {$\ket{\ell}$};
\node at (2.9,-2) {$\ket{r}$};

\end{tikzpicture}

\caption{Entanglement from the bending of light around a superposed mass. A heavy mass (shaded) is prepared in a spatial superposition $\ket{L} + \ket{R}$, and a light beam prepared in some initial wavepacket $\ket{0}$ is scattered. This produces an entangled state through evolution of the form $(\ket{L} + \ket{R}) \otimes \ket{0} \to \ket{L \ell} + \ket{R r}$.}
\label{figure-light}
\end{figure}
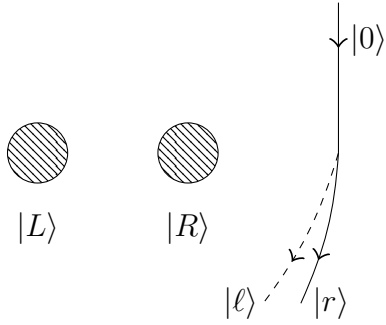

It is not too difficult, at least in principle, to extend these tabletop experiments to be spin-sensitive. Consider for example generating entanglement through the bending of light, as in Fig. \ref{figure-light}. In general relativity, the deflection angle of the light depends on both the $g_{00}$ and $g_{ij}$ components of the metric. In a scalar gravity model, one would typically get an incorrect deflection angle (e.g., the famous ``factor of 2'' difference with light bending in the $m \to 0$ limit of Newtonian gravity).\footnote{In the specific dilaton-type model considered here where $g_{\mu\nu} = \phi \eta_{\mu\nu}$, light actually doesn't bend at all, because it is coupled to gravity through $\mathcal{L} = \sqrt{-{\rm det}~g} g_{\alpha \beta} g_{\gamma \delta} F^{\alpha \gamma} F^{\beta \delta} = \eta_{\alpha \beta} \eta_{\gamma \delta} F^{\alpha \gamma} F^{\beta \delta}$. In other scalar gravity examples, for example Brans-Dicke gravity \cite{brans1961mach}, the answer may not be so simple but will still generically differ from general relativity.} More generally, different tensor structures in the interaction will give rise to different dependence on the momenta in the numerators of these scattering amplitudes, as seen in comparing $N_2$ with $N_0$.

Let us estimate the size of the entanglement in an experiment like that depicted in Fig. \ref{figure-light}. The deflection angle of the light is classically given by $\theta = G_N M/c^2 b$ where $b$ is the impact parameter and $M$ is the heavy source mass. Thus the differential bending along the two paths shown in Fig. \ref{figure-light} is of order
\be
\Delta \theta = \frac{G_N M}{c^2 b} \frac{\Delta b}{b} \approx 7.4 \times 10^{-27} \times \left( \frac{M}{1~{\rm g}} \right) \left( \frac{\Delta b}{10~{\rm \mu m}} \right) \left( \frac{100~{\rm \mu m}}{b} \right)^2 
\ee
where $\Delta b$ is the distance over which the source mass $M$ is superposed. This would be impossible to detect. One could improve the situation by having the light reflect back and forth within a cavity $N$ times. For a cavity of length $L$, the differential movement of the two forward peaks during one crossing would be $\Delta \ell = L \Delta \theta$. To get the two peaks separated by of order one laser wavelength $\lambda$ then requires $N \Delta \ell = N L \Delta \theta = \lambda$, meaning $N = \lambda/L \Delta \theta$, or a total integration time
\be
T = N \frac{L}{c} = \frac{b^2 \lambda c}{G_N M \Delta b} \approx 10^{16}~{\rm s}
\ee
with the same parameters given above. This can be improved by a factor $\sqrt{n_{\gamma}}$ with $n_{\gamma}$ photons in the cavity. Amusingly, allowing for a one second integration time, this works out to the requirement of $\sim 10^{32}$ laser photons, each of energy $(1000~{\rm nm})^{-1} = 0.2~{\rm eV}$. In terms of mass, this is $m = 10^{32} \times 0.2~{\rm eV} \approx 10^4~m_{\rm pl}$. 

Beyond the many practical difficulties \cite{Spengler:2021rlg}, the decay of this laser field into $e^+e^-$ pairs would render this fundamentally impossible \cite{sauter1931verhalten,schwinger1951gauge}. However, the fact that the required effective mass works out to be in the ballpark of $m_{\rm pl} \approx 0.2~{\rm \mu g}$ motivates some analogue of this experiment with two composite masses, much like the Newtonian experiments currently in development. One would need to engineer the coupling in a way sensitive to the tensor structure of the $h_{\mu\nu}T^{\mu\nu}$ interaction.

\section{Conclusions}

Experiments in the reasonably near future will be able to test whether or not gravity can entangle pairs of non-relativistic masses. Observation of this Newtonian entanglement would provide compelling but incomplete evidence for the existence of the graviton. These experiments would definitely rule out models where gravity emerges in something like a semi-classical sense. Such a model is a logical possibility consistent with all classical gravitational phenomena observed to date.

However, to make a definitive statement about the graviton itself, two key loopholes need to be closed. The simplest is that the Newtonian experiments are insensitive to the spin of the graviton. A more sophisticated experiment, like the one described in section \ref{section-spin}, could resolve this. 

The more difficult issue is that the specific unitarity assumptions used in this paper could be too strong. It is important to understand that we have imposed unitarity and Lorentz invariance as conditions on arbitrary scattering processes. It may be that an emergent gravitational interaction can still generate the appropriate entanglement signatures these experiments seek without admitting a unitary, Hamiltonian description like \eqref{Hnon-rel}. Construction of or no-go statements about such a model would be extremely valuable. But even if a graviton were not required in such a model, observation of Newtonian entanglement would teach us something remarkable about the detailed way in which gravity emerges: it would have to be able to coherently communicate quantum information between spatially disjoint systems.

\section*{Acknowledgements}

I thank Daniele Alves, Niklas Beisert, Nikita Blinov, Sougato Bose, Colby Delisle, Anson Hook, Matthew McCollough, Gordon Semenoff, David Simmons-Duffin, Raman Sundrum, Jacob Taylor, Oleksandr Tomalek, Jessica Turner, Mark van Raamsdonk, Junwu Wang, and Jordan Wilson-Gerow for discussions. I am especially grateful to Simon Caron-Huot for the suggestion to consider the six-point amplitudes of section \ref{section-tree}, and pointing me to reference \cite{dickinson2016probabilities}. I also gratefully acknowledge hospitality at the Cornell and University of Maryland physics departments while this work was being completed. My work at LBNL is supported by the US Department of Energy under contract DE-AC02-05CH11231 and by the Quantum Information Science Enabled Discovery (QuantISED) for High Energy Physics grant KA2401032.

\begin{appendix}

\section{Classical and semiclassical gravity models}
\label{appendix-classical}

In this appendix we briefly review the idea that gravity could be ``classical''. The quotes represent the fact that this could mean a variety of things in detail. There is a large literature on these ideas, see for example \cite{carney2019tabletop,Tilloy:2018tjp} for much more extensive discussion. Here I want to just highlight why it is possible to formulate some kind of classical gravity coupled to quantum matter in a way that can avoid the classic no-go statements of Weinberg and Polchinski \cite{weinberg1989testing,Polchinski:1990py}.

The naive starting point for a classical gravity model would be to source the Einstein equation with the expectation value of the stress tensor $G_{\mu\nu} = 8\pi G_N \braket{T_{\mu\nu}}$. This does not specify the full dynamics because we need an equation of motion for the matter state $\ket{\psi}$. Closing the system with a simple Schr\"{o}dinger equation leads to fundamental problems. The Newtonian limit is sufficient to see the basic issue. Imagine trying to define a model like\footnote{In this section, and only this section, operators are written with hats for clarity.}
\be
\label{semiclassical}
i \partial_t \ket{\psi} = \left[ \hat{H}_{\rm matter} + \sum_{i} V_i(\hat{\mb{x}}_i) \right] \ket{\psi}
\ee
with the indices $i,j = 1, \ldots, N$ running over some set of $N$ particles, $\hat{H}_{\rm matter}$ describing any non-gravitational evolution, and a ``semiclassical'' gravitational potential
\be
\label{semiclassical2}
V_i(\hat{\mb{x}}_i) = -\sum_{j \neq i} \frac{G_N m}{|\hat{\mb{x}}_i - \braket{\hat{\mb{x}}_j}|}.
\ee
In a limit where all the particle wavefunctions are strongly peaked on classical trajectories, we can approximate $\hat{\mb{x}}_i \approx \mb{x}_i$ as $c$-numbers, and the system \eqref{semiclassical} exactly reproduces classical Newton gravity. However, the interaction in general is a direct sum, and thus cannot generate entanglement. In this sense, the model is ``classical''.

The trouble is that this is a non-linear modification of the Schr\"{o}dinger dynamics, since $V$ depends quadratically on $\ket{\psi}$. It was noted long ago that generic such modifications lead to fundamental problems when applied to states where the classical approximation fails. For example, one can use Schr\"{o}dinger-cat type states to superluminally signal \cite{Polchinski:1990py,Yang:2012mh,bahrami2014role}.

\begin{figure}[t]
\centering 

\begin{tikzpicture}[scale=0.5]

\draw (0,3) -- (4,3);
\node at (-1.7,3) {$S_1: \ \ \ket{\psi_1}$};

\draw (0,1) -- (4,1);
\node at (-1.7,1) {$A_1: \ \ \ket{\phi}$};

\draw (0,-1) -- (4,-1);
\node at (-1.7,-1) {$A_2: \ \ \ket{\phi}$};

\draw (0,-3) -- (4,-3);
\node at (-1.7,-3) {$S_2: \ \ \ket{\psi_2}$};

\draw [black] (4,0.5) rectangle (6,3.5);
\node at (5,2) {$U_{ent}$};

\draw [black] (4,-0.5) rectangle (6,-3.5);
\node at (5,-2) {$U_{ent}$};

\draw (6,1) -- (8,1);
\draw [black] (8,1.5) -- (8,.5) arc(-90:90:.5) -- cycle;

\draw (6,-1) -- (8,-1);
\draw [black] (8,-1.5) -- (8,-.5) arc(90:-90:.5) -- cycle;

\draw (6,3) -- (11,3);
\draw [black] (11,2.5) rectangle (13,3.5);
\node at (12,3) {$U_{FB}$};

\draw (6,-3) -- (11,-3);
\draw [black] (11,-2.5) rectangle (13,-3.5);
\node at (12,-3) {$U_{FB}$};

\draw [double] (8.5,1) -- (11,-2.5);
\draw [double] (8.5,-1) -- (11,2.5);

\draw (13,3) -- (16,3);
\draw [dashed] (8.5,1) -- (16,1);

\draw (13,-3) -- (16,-3);
\draw [dashed] (8.5,-1) -- (16,-1);

\end{tikzpicture}

\caption{Adversarial measurement-and-feedback gravity, following \cite{Kafri:2014zsa}. The figure is taken from \cite{carney2019tabletop}.}
\label{figure-mfgravity}
\end{figure}
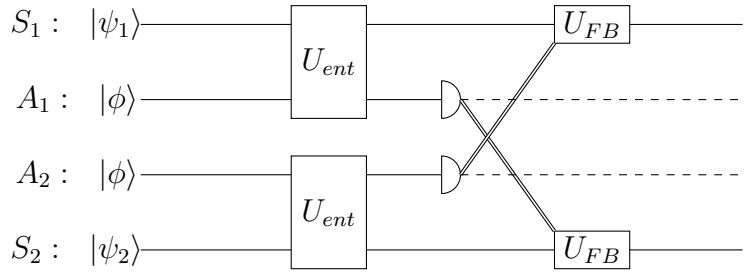

However, we can construct a model designed to evade these difficulties \cite{Kafri:2014zsa}. The basic strategy is to realize the non-linearity in the Schr\"{o}dinger equation as arising from a normal quantum mechanical process. We might call this approach ``adversarial gravity''. We imagine that the universe is a quantum simulation designed specifically to make it look like a classical gravitational force exists. See Fig. \ref{figure-mfgravity} for a circuit diagram of one timestep of the simulation, with two masses for simplicity. 

The computer begins by using some ancillae degrees of freedom to weakly measure the positions of the masses, allowing it to make estimates for $\braket{\hat{\mb{x}}_i}$ with $i=1,2$. The computer uses this information to compute a semiclassical Newton potential \eqref{semiclassical2}. This information is then used to evolve the state of the masses through a feedback unitary
\be
U_{\rm fb} = \exp \left\{ -i \sum_{i} V_i(\hat{\mb{x}}_i) dt \right\}.
\ee
It is clear that this unitary is a product $U = U_1 \otimes U_2$, so it cannot generate any entanglement between the two masses. This model produces a Schr\"{o}dinger-like evolution for the matter state of the form
\be
\label{mfgravity}
d \ket{\psi} = -i H_{\rm mat} dt \ket{\psi} - i \sum_{i} V_i(\hat{\mb{x}}_i) dt \ket{\psi} + \sum_i \sqrt{\gamma} \xi_i dW \ket{\psi} + \sum_{i,j} \xi_i \xi_j dt \ket{\psi}.
\ee
This is like \eqref{semiclassical}, except with the addition of the two final terms. These represent noise due to the weak measurement step. The differential $dW$ is a classical stochastic process satisfying the usual It\^{o} rule $dW^2 = dt$, and $\gamma,\xi$ represent the strength of the noise. We have replaced the deterministic equations \eqref{semiclassical}, \eqref{semiclassical2} with a stochastic evolution law. There are still non-linear terms in \eqref{mfgravity}, but their origin comes from quantum measurements via ancilla degrees of freedom which are traced out \cite{diosi1986universal}. Such non-linear terms are hardly pathological; they arise for example in any experiment which uses a measurement-and-feedback protocol. The description here was non-relativistic, but similar relativistic models of non-linear quantum mechanics also exist \cite{stamp2015rationale,barvinsky2018structure,Kaplan:2021qpv}.

This model is a kind of emergent gravity scenario. In this example the emergent gravitational force cannot entangle objects. It is not at all clear that this is a general property of emergent interactions. Indeed, one can construct models of entropic forces which produce quantum coherent interactions \cite{wang2016quantum}. Moreover, in AdS/CFT, the bulk graviton exists by construction, at least in the traditional sense of AdS/CFT as a string duality \cite{Maldacena:1997re}, since the closed string has a massless spin-2 excitation. This suggests an interesting question as to whether one can realize emergent gravity in the sense of Jacobson \cite{jacobson1995thermodynamics} or Verlinde \cite{Verlinde:2010hp} in such a way that gravity can communicate quantum information but remains a fundamentally open system.

\section{Details of amplitude calculations}
\label{appendix-details}

\subsection*{External probe interaction}

The $3 \to 3$ process of section \ref{section-tree} used an external probe system (a ``photon'') to model the effect the experimentalists interacting with the massive, gravitationally-coupled objects. Before moving to the detailed calculation of the $3 \to 3$ amplitude we thus need to provide some details about this interaction. This will also provide an excellent warmup to the $3 \to 3$ amplitude itself.

In our non-relativistic language, we will model the external probe as a massless scalar (i.e., a photon in the limit that we can ignore the polarizations). In abuse of language I will refer to these probe particles as photons throughout. The coupling is, in the interaction picture,
\be
\label{Ve}
V_e(t) = \lambda \int d^3 \mb{k} d^3\mb{p} d^3\mb{p}' \delta^3(\mb{k}+\mb{p}-\mb{p}')  \left[ W_{\mb{k}\mb{p},\mb{p}'} e^{i(E_{\mb{k}}+E_{\mb{p}} - E_{\mb{p}'})t} \ket{\mb{k} \mb{p}} \bra{\mb{p}'} + {\rm h.c.} \right].
\ee
The coupling $\lambda$ has mass dimension one and the matrix elements $W$ are given by
\be
\label{Welements}
W_{\mb{k} \mb{p},\mb{p}'} = \frac{1}{\sqrt{[2 (2\pi)^3]^3 E_{\mb{k}} E_{\mb{p}} E_{\mb{p}'}}},
\ee
thus carrying dimension ${\rm mass}^{-3/2}$. I will use $\mb{k}$ for the probe momentum and $\mb{p},\mb{p}'$ for the massive object. The form of this matrix element can be derived for example by considering the massive object as a scalar particle and quantizing the interaction $V_e = \lambda \int d^3\mb{x} \chi^2 \phi$ with $\phi$ the probe and $\chi$ the massive object. 

\begin{figure}[t]
\centering
\begin{tikzpicture}[yscale=.8]

\draw (-2,-1.5) -- (-1,-.5);
\draw [photon] (0,-1.5) -- (-1,-.5);
\draw (-1,-.5) -- (-1,.5);
\draw (-1,.5) -- (-2,1.5);
\draw [photon] (-1,.5) -- (0,1.5);

\node at (-2,-1.8) {$\mb{p}$};
\node at (0,-1.8) {$\mb{k}$};
\node at (-2,1.8) {$\mb{p}'$};
\node at (0,1.8) {$\mb{k}'$};

\begin{scope}[xshift=140]
\draw (-2,-1.5) -- (-1,-.5);
\draw [photon] (0,1.5) -- (-1,-.5);
\draw (-1,-.5) -- (-1,.5);
\draw (-1,.5) -- (-2,1.5);
\draw [photon] (-1,.5) -- (0,-1.5);

\node at (-2,-1.8) {$\mb{p}$};
\node at (0,-1.8) {$\mb{k}$};
\node at (-2,1.8) {$\mb{p}'$};
\node at (0,1.8) {$\mb{k}'$};
\end{scope}

\end{tikzpicture}
\caption{Diagrams for the probe-system interactions, one for each term in \eqref{Torder-compton}.}
\label{figure-compton}
\end{figure}
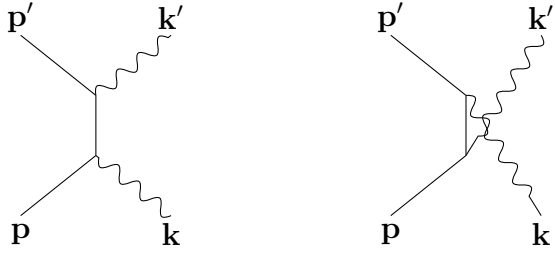

To get a feel for this interaction, consider a process with a single mass which absorbs and emits a probe photon. This means we need to go to second order in the Dyson series
\be
S^{(2)} = \frac{(-i)^2}{2!} \int_{-\infty}^{\infty} dt_1 dt_2 \braket{\beta | \mathcal{T} V_e(t_1) V_e(t_2) | \alpha},
\ee
with initial state $\ket{\alpha} = \ket{\mb{k} \mb{p}}$ and final state $\ket{\beta} = \ket{\mb{k}' \mb{p}'}$. There are two contributions to this due to the time-ordering
\be
\label{Torder-compton}
\mathcal{T} V_e(t_1) V_e(t_2) = V_e(t_1) V_e(t_2) \Theta(t_1 - t_2) + V_e(t_2) V_e(t_1) \Theta(t_2 - t_1).
\ee
Each of these terms will further contribute two terms, one for each ordering of the emission and absorption [i.e. one each for either the $W$ or $W^*$ term in \eqref{Ve}]. Using the explicit matrix element \eqref{Ve}, the two terms which contribute to the first diagram in Fig. \ref{figure-compton}, representing absorption followed by emission, are
\begin{align}
\begin{split}
S_{A} & = -\frac{\lambda^2}{2} \int_{-\infty}^{\infty} dt_1 dt_2 d^3\mb{q} W_{\mb{k}'\mb{p}',\mb{q}} e^{i(E_{\mb{k}'}+E_{\mb{p}'} - E_{\mb{q}})t_2} \delta^3(\mb{k}' + \mb{p}' - \mb{q}) \\
& \times W^*_{\mb{q},\mb{k}\mb{p}} e^{i(E_{\mb{q}}-E_{\mb{k}} - E_{\mb{p}})t_1} \delta^3(\mb{q} - \mb{k} - \mb{p}) \Theta(t_2-t_1) \\
& + (t_1 \leftrightarrow t_2),
\end{split}
\end{align}
where the last line just means a copy of the previous term but with the time variables switched. To do the time integrals, we can re-write the step function using the identity 
\be
\label{step}
\Theta(t) e^{-i E t} = \frac{1}{2\pi i} \int_{-\infty}^{\infty} d\omega \frac{e^{i \omega t}}{\omega - E + i \epsilon}
\ee
which can be easily checked by contour integration. Here and throughout, the limit $\epsilon \to 0$ is taken at the end of all computations. Using this on the phase $e^{-i E_{\mb{q}}(t_2 - t_1)}$ and evaluating the time integrals we obtain
\begin{align}
\begin{split}
\label{S12}
S_{A} & = -\frac{2 (2\pi)^2\lambda^2}{4\pi i} \int d\omega d^3\mb{q} W_{\mb{k}'\mb{p}',\mb{q}} \delta(E_{\mb{k}'}+E_{\mb{p}'} - \omega) \delta^3(\mb{k}' + \mb{p}' - \mb{q}) \\
& \times \frac{1}{\omega - E_{\mb{q}} + i \epsilon} W^*_{\mb{q},\mb{k}\mb{p}} \delta(\omega-E_{\mb{k}} - E_{\mb{p}}) \delta^3(\mb{q} - \mb{k} - \mb{p}).
\end{split}
\end{align}
The overall factor of 2 appears because the $(t_1 \leftrightarrow t_2)$ term contributes identically. The physical interpretation of the $i \epsilon$ here can be understood from \eqref{step}. This says that the intermediate positive-energy state of energy $E_{\mb{q}}$ has energy flowing forward in time.

Now we would like to apply the Lorentzian bootstrap idea to this expression. In other words, we want this to be a Lorentz-invariant function times the overall wavefunction re-scaling $\sim 1/\sqrt{E_{\mb{p}} E_{\mb{k}} E_{\mb{p}'} E_{\mb{k}'}}$ as in \eqref{M1rel}. This re-scaling is already taken care of in the matrix elements \eqref{Welements}, which also produce an additional factor $1/E_{\mb{q}}$. Identifying $\omega = q^0$, the delta-functions can be combined in the obvious way to produce a pair of four-dimensional delta functions. 

The tricky part is the non-relativistic propagator $\sim 1/(\omega - E_{\mb{q}} + i \epsilon)$. To get something Lorentz-invariant we will need to make a quadratic function out of the denominator, and the only option is to use $q^2$. The function $q^2 + m^2$ has precisely the same pole $\omega = q^0 = + E_{\mb{q}}$ (once we move to the relativistic dispersion relation $E_{\mb{q}} = \sqrt{m^2 + \mb{q}^2}$). Specifically, we have
\be
\label{poles}
\frac{1}{q^2+m^2} = \frac{-1}{2 E_{\mb{q}}} \left[ \frac{1}{q^0 - E_{\mb{q}}} - \frac{1}{q^0 + E_{\mb{q}}} \right].
\ee
The remaining question is the prescription for handling integration of the two poles $q^0 = \pm E_{\mb{q}}$. In the non-relativistic propagator above, we saw that the positive-energy pole was displaced $E_{\mb{q}} \to E_{\mb{q}} - i \epsilon$, so the only issue is what to do with this new negative-energy pole. We will see in the next section that the only answer consistent with Lorentz invariance and the optical theorem is to displace this pole $-E_{\mb{q}} \to -(E_{\mb{q}} - i \epsilon)$.\footnote{This is not quite true. What is inconsistent would be to use the retarded or advanced prescriptions, in which both poles are shifted above the real axis or both are shifted below the real axis. One could, however, use the prescription $q^2 + m^2 + i\epsilon$ instead of the usual Feynman $q^2 + m^2 -i \epsilon$. This would correspond to an overall time-reversal, i.e., a model in which the definition of positive and negative energies are reversed. One can make interesting models in which both conventions are used \cite{kaplan2006symmetry,donoghue2019arrow}. In what follows we will use the standard Feynman prescription.}  We will thus identify
\be
\label{energyprop}
\frac{1}{\omega - E_{\mb{q}} + i \epsilon} \to -\frac{2 E_{\mb{q}}}{q^2 + m^2 - i \epsilon}.
\ee
This is not Lorentz-invariant, but the numerator cancels the extra factor of $1/E_{\mb{q}}$ we picked up from the interaction vertices, leaving an overall Lorentz-invariant expression. 

With this identification, we can do the integral in \eqref{S12} using one set of the delta-functions, and obtain
\be
S_{A} = \frac{1}{\sqrt{[2 (2\pi)^3]^4 E_{\mb{k}} E_{\mb{p}} E_{\mb{k}'} E_{\mb{p}'}}} \frac{i \lambda^2}{(2\pi)^2} \frac{1}{(p+k)^2 + m^2 - i \epsilon} \delta^4(p + k - p' - k').
\ee
By the exact same logic, the other two terms (absorption followed by emission, the second part of Fig. \ref{figure-compton}) produce a contribution $S_{B}$ identical to $S_{A}$ except  with the denominator $(p + k)^2 \to (p - k')^2$. In terms of our Feynman amplitudes, this means we would identify 
\be
M_{\mb{k} \mb{p} \to \mb{k}' \mb{p}'} =  \frac{\lambda^2}{(2\pi)^3} \left[ \frac{1}{(p+k)^2 + m^2 - i \epsilon} + \frac{1}{(p-k')^2 + m^2 - i \epsilon} \right].
\ee
This argument contains all the same pieces we will need to evaluate the $3 \to 3$ amplitude of section \ref{section-tree}, to which we now turn.

\subsection*{Six-point amplitude derivation}

Consider the process involving two gravitationally coupled masses $m_1, m_2$ and some external ``photons'' depicted in Fig. \ref{figure-6pt}. For simplicity we will assume that there are two distinguishable probes,
\be
\label{Vi}
V_{i}(t) = \lambda \int d^3 \mb{k} d^3\mb{p}_i d^3\mb{p}_i' \delta^3(\mb{k}+\mb{p}_i-\mb{p}_i')  \left[ W_{\mb{k}\mb{p}_i,\mb{p}_i'} e^{i(E_{\mb{k}}+E_{\mb{p}_i} - E_{\mb{p}_i'})t} \ket{\mb{k} \mb{p}_i} \bra{\mb{p}_i'} + {\rm h.c.} \right] \otimes 1_{j\neq i}
\ee
which will simplify some combinatoric factors.

\begin{figure}[t]
\centering

\begin{tikzpicture}[scale=.7,every node/.style={scale=0.7}]

\draw (-2,-1) -- (-1,0);
\draw (-2,1) -- (-1,0);
\draw [fill=lightgray] (0,0) ellipse [x radius=1, y radius=.1];
\draw (1,0) -- (2,1);
\draw (1,0) -- (2,-1);
\draw [photon] (-1.5,-2) -- (-1.5,-.5);
\draw [photon] (1.5,2) -- (1.5,.5);

\node at (0,.4) {$V_N(t)$};

\node at (-2,1.3) {$\mb{p}'_1$};
\node at (2,1.3) {$\mb{p}'_2$};
\node at (-2,-1.3) {$\mb{p}_1$};
\node at (2,-1.3) {$\mb{p}_2$};
\node at (-1.2,-2) {$\mb{k}$};
\node at (1.2,2) {$\mb{k}'$};

\node at (2.8,0) {$+$};

\begin{scope}[xshift=160]

 \draw (-2,-1) -- (-1,0);
\draw (-2,1) -- (-1,0);
\draw [fill=lightgray] (0,0) ellipse [x radius=1, y radius=.1];
\draw (1,0) -- (2,1);
\draw (1,0) -- (2,-1);
\draw [photon] (-1.5,-2) -- (-1.5,-.5);
\draw [photon] (1.5,-2) -- (1.5,-.5);

\node at (2.8,0) {$+$};

\end{scope}

\begin{scope}[xshift=320]

 \draw (-2,-1) -- (-1,0);
\draw (-2,1) -- (-1,0);
\draw [fill=lightgray] (0,0) ellipse [x radius=1, y radius=.1];
\draw (1,0) -- (2,1);
\draw (1,0) -- (2,-1);
\draw [photon] (-1.5,2) -- (-1.5,.5);
\draw [photon] (1.5,-2) -- (1.5,-.5);

\node at (2.8,0) {$+$};

\end{scope}

\begin{scope}[xshift=480]

 \draw (-2,-1) -- (-1,0);
\draw (-2,1) -- (-1,0);
\draw [fill=lightgray] (0,0) ellipse [x radius=1, y radius=.1];
\draw (1,0) -- (2,1);
\draw (1,0) -- (2,-1);
\draw [photon] (-1.5,2) -- (-1.5,.5);
\draw [photon] (1.5,2) -- (1.5,.5);

\end{scope}

\end{tikzpicture}

\caption{Contributions to the scattering amplitude for the $3 \to 3$ process with external photons.}
\label{figure-6pt-all}

\end{figure}
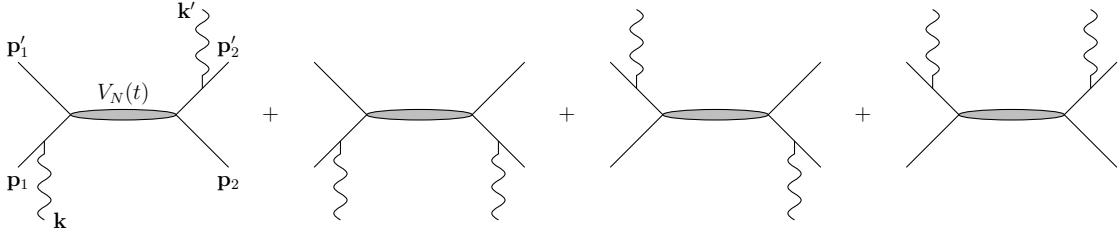

Starting from the Dyson series \eqref{dyson} we need to go to third order in the couplings. More specifically we need to go to order $\lambda^2 G_N$ and consider the $S$-matrix contribution
\be
S^{(3)} = \int_{-\infty}^{\infty} dt_1 dt_2 dt_N \braket{\beta | \mathcal{T} V_1(t_1) V_2(t_2) V_N(t_N) | \alpha}.
\ee
This gives six independent terms from the time-ordering symbol
\begin{align}
\mathcal{T} V_1 V_2 V_N = V_1 V_N V_2 \Theta(t_1 - t_N) \Theta(t_N - t_2) + V_2 V_N V_1 \Theta(t_2 - t_N) \Theta(t_N - t_1) + \cdots, 
\end{align} 
which can be represented by a sum of four Feynman diagrams, as in Fig. \ref{figure-6pt-all}. We will now proceed to compute one of these in detail in terms of our Lorentzian bootstrap. After we are done it will be clear that the others just add essentially identical factors. Alternatively, one could integrate against localized-in-time wavepackets for the probes, which would suppress the contributions from the other diagrams.

Let's study a particular term corresponding to the first diagram of Fig. \ref{figure-6pt-all}, namely the $S$-matrix contribution
\be
S_{1N2} = \int_{-\infty}^{\infty} dt_1 dt_2 dt_N \braket{\beta | V_2(t_2) V_N(t_N) V_1(t_1) | \alpha} \Theta(t_2 - t_N) \Theta(t_N - t_1).
\ee
Inserting the explicit forms of the interactions \eqref{VN-int}, \eqref{Vi} and using the states $\ket{\alpha} = \ket{\mb{k} \mb{p}_1 \mb{p}_2}$, $\ket{\beta} = \ket{\mb{k}' \mb{p}_1' \mb{p}_2'}$, we have
\begin{align}
\begin{split}
S_{1N2} & = \lambda^2 \int_{-\infty}^{\infty} dt_2 dt_N dt_1 d^3\ul{\mb{p}}_1 d^3\ul{\mb{p}}_2 d^3 \ul{\ul{\mb{p}}}_1 d^3 \ul{\ul{\mb{p}}}_2 \Theta(t_2 - t_N) \Theta(t_N - t_1) \\
& \times W_{\mb{k}'\mb{p}_2',\ul{\ul{\mb{p}}}_2} e^{i(E_{\mb{k}'}+E_{\mb{p}_2'} - E_{\ul{\ul{\mb{p}}}_2})t_2} \delta^3(\mb{k}' + \mb{p}_2' - \ul{\ul{\mb{p}}}_2) \delta^3(\mb{p}_1' - \ul{\ul{\mb{p}}}_1) \\
& \times V_{\ul{\ul{\mb{p}}}_1\ul{\ul{\mb{p}}}_2, \ul{\mb{p}}_1 \ul{\mb{p}}_2}  e^{i (E_{\ul{\ul{\mb{p}}}_2} + E_{\ul{\ul{\mb{p}}}_1} - E_{\ul{\mb{p}}_2} - E_{\ul{\mb{p}}_1})t_N} \delta^3(\ul{\ul{\mb{p}}}_1 + \ul{\ul{\mb{p}}}_2 - \ul{\mb{p}}_1 - \ul{\mb{p}}_2) \\
& \times W_{\ul{\mb{p}}_1,\mb{k} \mb{p}_1} e^{i(E_{\ul{\mb{p}}_1} - E_{\mb{k}} - E_{\mb{p}_1}) t_1} \delta^3(\ul{\mb{p}}_1 - \mb{k} - \mb{p}_1) \delta^3(\ul{\mb{p}}_2 - \mb{p}_2). 
\end{split}
\end{align}
Using the step function identity \eqref{step} in much the same way as before, the time integrals may be evaluated in terms of some frequency integrals,
\begin{align}
\begin{split}
& \int_{-\infty}^{\infty} dt_2 dt_N dt_1 \Theta(t_2 - t_N) \Theta(t_N - t_1) \\
& \times e^{i(E_{\mb{k}'}+E_{\mb{p}_2'} - E_{\ul{\ul{\mb{p}}}_2})t_2} e^{i (E_{\ul{\ul{\mb{p}}}_2} + E_{\ul{\ul{\mb{p}}}_1} - E_{\ul{\mb{p}}_2} - E_{\ul{\mb{p}}_1})t_N} e^{i(E_{\ul{\mb{p}}_1} - E_{\mb{k}} - E_{\mb{p}_1}) t_1} \\
& = \int \frac{d\omega_2}{\omega_2 - E_{\ul{\ul{p}}_2} + i \epsilon}  \frac{d\omega_1}{\omega_1 - E_{\ul{p}_1} + i \epsilon} \\
& \times \delta(E_{\mb{k}'} + E_{\mb{p}_2'} - \omega_2) \delta(E_{\ul{\ul{\mb{p}}}_1} + \omega_2 - \omega_1 - E_{\ul{\mb{p}}_2} ) \delta(\omega_1 - E_{\mb{p}_1} - E_{\mb{k}} ).
\end{split}
\end{align}
Performing the $d^3\ul{\ul{\mb{p}}}_1 d^3\ul{\mb{p}}_2$ momentum integrals with the simple delta functions then leaves free integrals over $d\omega_1 d\omega_2 d^3\ul{\mb{p}}_1 d^3\ul{\ul{\mb{p}}}_2$. Identifying $\omega_1 = \ul{p}_1^0, \omega_2 = \ul{\ul{p}}_2^0$ and forming four-vectors we then obtain
\begin{align}
\begin{split}
S_{1N2} & = \lambda^2 \int d^4\ul{p}_1 d^4\ul{\ul{p}}_2 W_{\mb{k}'\mb{p}_2',\ul{\ul{\mb{p}}}_2} \delta^4(k' + p_2' - \ul{\ul{p}}_2) \frac{1}{\omega_2 - E_{\ul{\ul{p}}_2} + i \epsilon} V_{\mb{p}_1' \ul{\ul{\mb{p}}}_2, \ul{\mb{p}}_1 \mb{p}_2}  \\
& \times \delta^4(p_1' + \ul{\ul{p}}_2 - \ul{p}_1 - p_2) \frac{1}{\omega_1 - E_{\ul{\mb{p}}_1} + i \epsilon} W_{\ul{\mb{p}}_1,\mb{k} \mb{p}_1} \delta^4(\ul{p}_1 - k - p_1) .
\end{split}
\end{align}
Again, everything in the above is simply a formal manipulation of the non-relativistic amplitude. 

Finally, we now make the same ``bootstrap'' identifications as in earlier sections. The energy denominators are replaced according to \eqref{energyprop}, while the Newton interaction matrix elements are replaced in line with the discussion of section \ref{section-bootstrap}. We include the various $1/\sqrt{E}$ factors appropriately. Performing the two remaining momentum integrals, we then identify the resulting $S$-matrix contribution:
\begin{align}
\begin{split}
S_{1N2} & \to \frac{\lambda^2 G_N m^4}{\sqrt{[2 (2\pi)^{3}]^6 E_{\mb{k}} E_{\mb{p}_1} E_{\mb{p}_2} E_{\mb{k}'} E_{\mb{p}_1'} E_{\mb{p}_2'}}} \delta^4(k' + p_1' + p_2' - k - p_1 - p_2) \\
& \times \left( \frac{1}{(p_1 + k)^2 + m^2 - i \epsilon}\right) \left(  \frac{1}{\tilde{k}^2 + \mu^2 - i \epsilon} \right) \left( \frac{1}{(p_2' + k')^2 + m^2 - i \epsilon} \right),
\end{split}
\end{align}
where $\tilde{k} = p_1' - (p_1 + k)$. Comparing to our $S$-matrix conventions \eqref{feynman} then leads to the identification of the Feynman amplitude
\begin{align}
M = \lambda^2 G_N m^4 \left( \frac{1}{(p_1 + k)^2 + m^2 - i \epsilon}\right) \left(  \frac{1}{\tilde{k}^2 + \mu^2 - i \epsilon} \right) \left( \frac{1}{(p_2' + k')^2 + m^2 - i \epsilon} \right),
\end{align}
precisely as in \eqref{M-tree}.

At this stage, we can close the loop and explain the need for the specific $i\epsilon$ pole prescription used here. In the previous section, I claimed that this (or its $-i\epsilon$ time-reversed version) is the only prescription consistent with Lorentz invariance and unitarity. To see this, consider what would happen if we used instead a prescription like
\be
\label{poles-ret}
\frac{1}{q^0 - E_{\mb{q}} - i \epsilon} - \frac{1}{q^0 + E_{\mb{q}} - i \epsilon},
\ee
in the notation of \eqref{poles}, in which both poles are shifted above the real $q^0$ axis. When we compute the left-hand side of the optical theorem \eqref{optical}, we would pick up a term proportional to the imaginary part of this factor, which is $\delta(q^0 - E_{\mb{q}}) - \delta(q^0 + E_{\mb{q}})$. This is not Lorentz invariant. In contrast, the usual Feynman prescription produces an imaginary part proportional to $\delta(q^0 - E_{\mb{q}}) + \delta(q^0 + E_{\mb{q}}) \propto \delta(q^2 + m^2)$, as used in the main text. Thus the retarded prescription \eqref{poles-ret} or the advanced prescription where both poles are shifted below the real axis would be incapable of satisfying the optical theorem \eqref{optical}, since the right-hand side is manifestly Lorentz-invariant.

\subsection*{Optical theorem for the six-point amplitude}

Finally, we spell out the details of the new amplitudes on the right-hand side of \eqref{opticalfinal}. Everything is almost identical to the $2 \to 2$ process of the previous subsection. The only change is the $\sqrt{G_N} m$ coupling, and an additional factor for the disconnected part:
\begin{equation}
\label{graviton-amplitude}
M_{\mb{k} \mb{p}_1 \mb{p}_2 \to \ul{\mb{k}} \ul{\mb{p}}_1 \ul{\mb{p}}_2} = 
\vcenter{\hbox{\begin{tikzpicture}[scale=.7,every node/.style={scale=.7}]
\draw (-2,-1.5) -- (-1,-.5);
\draw [photon] (0,-1.5) -- (-1,-.5);
\draw (-1,-.5) -- (-1,.5);
\draw (-1,.5) -- (-2,1.5);
\draw [dashed] (-1,.5) -- (0,1.5);

\draw (1,-1.5) -- (1,1.5);

\node at (-2,1.8) {$\ul{\mb{p}}_1$};
\node at (1,1.8) {$\ul{\mb{p}}_2$};
\node at (-2,-1.8) {$\mb{p}_1$};
\node at (1,-1.8) {$\mb{p}_2$};
\node at (0,-1.8) {$\mb{k}$};
\node at (0,1.8) {$\ul{\mb{k}}$};
\end{tikzpicture}}}
 = \frac{\sqrt{G_N} m^2 \lambda}{(p_1+k)^2 + m^2 - i \epsilon} \delta^3(\ul{\mb{p}}_2 - \mb{p}_2) \times 2 E_{\mb{p}_2} (2\pi)^3.
\end{equation}
The final factor $2 E_{\mb{p}_2} (2\pi)^3$ is needed to cancel out the Lorentzian phase-space factors of equation \eqref{feynman} for the disconnected particle. To apply this amplitude in the optical theorem \eqref{optical} to reproduce the unitarity condition \eqref{opticalfinal}, the non-trivial part of the sum over $\ket{X}$ (the sum over final three-body states) can be written explicitly
\begin{equation}
\label{sumX-tree}
\sum_{\ket{X}} B_X^2 \delta^4(p_X - p_{\alpha}) = \int \frac{d^3\ul{\mb{k}} d^3\ul{\mb{p}}_1 d^3\ul{\mb{p}}_2}{[2(2\pi)^3]^3 E_{\ul{\mb{k}}} E_{\ul{\mb{p}}_1} E_{\ul{\mb{p}}_2}} \delta^4(\ul{k} + \ul{p}_1 + \ul{p}_2 - k - p_1 - p_2),
\end{equation}
where $\ket{\alpha} = \ket{\mb{k} \mb{p}_1 \mb{p}_2}$. Using \eqref{graviton-amplitude} and \eqref{sumX-tree} on the right-hand side of \eqref{optical}, with $\ket{\beta} = \ket{\mb{k}' \mb{p}'_1 \mb{p}'_2}$, one finds
\begin{align}
\begin{split}
& \sum_{\ket{X}} B_X^2 \delta^4(p_X - p_{\alpha}) M_{\alpha \to X} M^*_{\beta \to X} = \int \frac{d^3\ul{\mb{k}} d^3\ul{\mb{p}}_1 d^3\ul{\mb{p}}_2}{[2(2\pi)^3]^3 E_{\ul{\mb{k}}} E_{\ul{\mb{p}}_1} E_{\ul{\mb{p}}_2}} \delta^4(\ul{k} + \ul{p}_1 + \ul{p}_2 - k - p_1 - p_2) \\
& \times \left[ \frac{\sqrt{G_N} m^2 \lambda}{(p_1+k)^2 + m^2 - i \epsilon} \delta^3(\ul{\mb{p}}_2 - \mb{p}_2) \times 2 E_{\ul{\mb{p}}_2} (2\pi)^3 \right] \\
& \times \left[ \frac{\sqrt{G_N} m^2 \lambda}{(p'_2+k')^2 + m^2 - i \epsilon} \delta^3(\ul{\mb{p}}_1 - \mb{p}'_1) \times 2 E_{\ul{\mb{p}}_1} (2\pi)^3 \right] \\
& = G_N m^4 \lambda^2 \int \frac{d^3\ul{\mb{k}}}{[2(2\pi)^3] E_{\ul{\mb{k}}}} \delta^4(\ul{k} + p'_1 - k - p_1) \frac{1}{(p_1+k)^2 + m^2 - i \epsilon} \frac{1}{(p'_2+k')^2 + m^2 - i \epsilon} \\
& = G_N m^4 \lambda^2 \delta(\ul{k}^2 + \mu^2) \frac{1}{(p_1+k)^2 + m^2 - i \epsilon} \frac{1}{(p'_2+k')^2 + m^2 - i \epsilon} \Big|_{\ul{k} = k + p_1 -  p'_1} \\
\end{split}
\end{align}
where we used the usual Lorentz phase space identity $\int d^3\mb{k}/((2\pi)^3 2E_{\mb{k}}) = \int d^4 k \delta^2(k^2+\mu^2) \Theta(k^0)$ before doing the integral to obtain the last line. Identifying $\ul{k} = \tilde{k}$ in the notation of the main text, this verifies the optical theorem \eqref{optical}, as described in \eqref{opticalfinal}. The sum over final states has collapsed to a discrete point in phase space, namely the specific final state $\ket{X} = \ket{\tilde{\mb{k}} \mb{p}_1 \mb{p}'_2}$ with a radiated ``graviton''.

\subsection*{Graviton and scalar graviton Feynman rules}

I record here the Feynman rules used in section \ref{section-implications}. We will use a scalar field for the matter for simplicity. In transverse-traceless gauge, the graviton-matter interactions lead to the Feynman rules \cite{Donoghue:1994dn,Donoghue:1995cz,Burgess:2003jk}
\begin{align}
\begin{split}
\vcenter{\hbox{\begin{tikzpicture}
\draw (-1,-1) -- (0,0);
\node at (-1,-1.3) {$p$};
\draw (-1,1) -- (0,0);
\node at (-1,1.3) {$p'$};
\draw [gluon] (0,0) -- (1.5,0);
\node at (1.5,.3) {$\alpha \beta$};
\end{tikzpicture}}} & = \sqrt{8\pi G_N} \left[ p_{\alpha} p'_{\beta} + p'_{\alpha} p_{\beta} - \eta_{\alpha \beta} (p \cdot p' + m^2) \right] \\
\vcenter{\hbox{\begin{tikzpicture}
\draw [gluon] (-1,0) -- (1,0);
\node at (-1,.3) {$\alpha \beta$};
\node at (1,.3) {$\gamma \delta$};
\node at (0,-.3) {$q$};
\end{tikzpicture}}} & = \frac{i}{q^2 - i \epsilon} \left[ \eta^{\alpha \gamma} \eta^{\beta \delta} + \eta^{\alpha \delta} \eta^{\beta \gamma} - \eta^{\alpha \beta} \eta^{\gamma \delta} \right].
\label{feynmanrules-graviton}
\end{split}
\end{align}
The Feynman rules for the scalar graviton model used in the main text can be obtained by simply tracing over the Lorentz indices in the spin-2 model. Specifically, one has
\begin{align}
\begin{split}
\vcenter{\hbox{\begin{tikzpicture}
\draw (-1,-1) -- (0,0);
\node at (-1,-1.3) {$p$};
\draw (-1,1) -- (0,0);
\node at (-1,1.3) {$p'$};
\draw [dashed] (0,0) -- (1.5,0);
\end{tikzpicture}}}  & = -2 \sqrt{8\pi G_N} \left[ p \cdot p' + 2 m^2 \right] \\
\vcenter{\hbox{\begin{tikzpicture}
\draw [dashed] (-1,0) -- (1,0);
\node at (0,-.3) {$q$};
\node at (0,.3) {};
\end{tikzpicture}}} & = \frac{i}{q^2 - i \epsilon}.
\end{split}
\end{align}

\section{Unitarity violation in particle-antiparticle scattering}
\label{appendix-particleantiparticle}

In the example of section \ref{section-main}, we stayed very close to the basic premise of the actual experiments. The massive ``particles'' there can be viewed directly as the massive, composite objects used to demonstrate Newtonian entanglement in concrete experimental proposals. However, to emphasize the generality of this argument, we now turn to a more sophisticated example, in which we use the crossing symmetry property of relativistic $S$-matrices to study gravitational particle-antiparticle scattering.

We thus add another postulate to our $S$-matrix rules:
\begin{enumerate}[label=(\Alph*)]
\setcounter{enumi}{2}
\item {\bf Crossing symmetry}. The matrix element for a process $A + B \to C + D$ is given by the same function as that for a process, like $A + B + \bar{C} \to D$, where an incoming/outgoing particle is exchanged with an outgoing/incoming antiparticle (and momenta switched accordingly).
\end{enumerate}
For example, the amplitude for electron-electron scattering $e^- e^- \to e^- e^-$ determines the amplitude for $e^- e^+ \to e^- e^+$. In quantum field theory, crossing is a consequence of micro-causality, the requirement that field operators satisfy $[\phi(x),\phi(x')] = 0$ for $x,x'$ space-like separated \cite{gell1954use}. It is rigorously satisfied in $2 \to 2$ scattering \cite{bros1965proof,mizera2021bounds}. The same property holds in string theory despite the absence of microscopic local fields \cite{mizera2021bounds}. Following the analytic $S$-matrix program \cite{eden2002analytic} we thus will simply assume it as a postulate of scattering processes.

\begin{figure}[t]

\centering

\begin{tikzpicture}[scale=0.8, every node/.style={transform shape}]

\draw [->-=.5] (-1.5,-2) -- (-1,-1);
\draw [->-=.5] (1.5,-2) -- (1,-1);
\draw [fill=lightgray] (0,-1) ellipse [x radius=1, y radius=.1];
\draw [->-=.5] (-1,-1) -- (-1,1);
\draw [->-=.5] (1,-1) -- (1,1);
\draw [fill=lightgray] (0,1) ellipse [x radius=1, y radius=.1];
\draw [->-=.5] (1,1) -- (1.5,2);
\draw [->-=.5] (-1,1) -- (-1.5,2);

\node at (0,-1.4) {$V_N(t_1)$};
\node at (0,1.4) {$V_N(t_2)$};
\node at (-1.4,0) {$\mb{k}_1$};
\node at (1.4,0) {$\mb{k}_2$};

\node at (-1.5,2.3) {$\mb{p}'_1$};
\node at (1.5,2.3) {$\mb{p}'_2$};
\node at (-1.5,-2.3) {$\mb{p}_1$};
\node at (1.5,-2.3) {$\mb{p}_2$};

\draw [thick,->] (2.2,0) -- (3.2,0);
\node at (2.7,.4) {Lorentz};

\begin{scope}[xshift=160]

\draw [->-=.5] (-1.5,-2) -- (-1,-1);
\draw [->-=.5] (1.5,-2) -- (1,-1);
\draw [dashed] (-1,-1) -- (1,-1);
\draw [->-=.5] (-1,-1) -- (-1,1);
\draw [->-=.5] (1,-1) -- (1,1);
\draw [dashed] (-1,1) -- (1,1);
\draw [->-=.5] (1,1) -- (1.5,2);
\draw [->-=.5] (-1,1) -- (-1.5,2);

\node at (-1.4,0) {$k_1$};
\node at (1.4,0) {$k_2$};

\node at (-1.5,2.3) {$\mb{p}'_1$};
\node at (1.5,2.3) {$\mb{p}'_2$};
\node at (-1.5,-2.3) {$\mb{p}_1$};
\node at (1.5,-2.3) {$\mb{p}_2$};

\draw [thick,->] (2.2,0) -- (3.2,0);
\node at (2.7,.4) {crossing};

\end{scope}

\begin{scope}[xshift=320]

\draw [->-=.5] (-1.5,-2) -- (-1,-1);
\draw  [-<-=.5] (1.5,-2) -- (1,-1);
\draw [->-=.5] (-1,-1) -- (1,-1);
\draw [dashed] (-1,-1) -- (-1,1);
\draw [dashed] (1,-1) -- (1,1);
\draw [-<-=.5] (-1,1) -- (1,1);
\draw [-<-=.5] (1,1) -- (1.5,2);
\draw [->-=.5] (-1,1) -- (-1.5,2);

\node at (-1.4,0) {$k_1$};
\node at (1.4,0) {$k_2$};

\node at (-1.5,2.3) {$\mb{p}'_1$};
\node at (1.5,2.3) {$\mb{p}'_2$};
\node at (-1.5,-2.3) {$\mb{p}_1$};
\node at (1.5,-2.3) {$\mb{p}_2$};

\end{scope}

\end{tikzpicture}

\caption{Basic procedure to compute the elastic $\chi \bar{\chi}$ scattering amplitude. The left diagram represents the non-relativistic $\chi \chi \to \chi \chi$ scattering ampitude, computed to second order in perturbation theory. Its relativistic extension is depicted in the middle panel. Finally, by crossing symmetry, the same mathematical expression can be used to compute the $\chi \bar{\chi} \to \chi \bar{\chi}$ amplitude to the same order, depicted in the third panel.}

\label{figure-chichibar}
\end{figure}
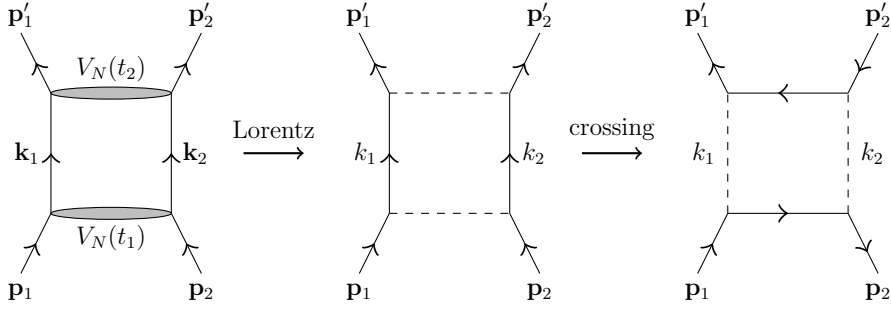

Consider the $2 \to 2$, purely gravitational scattering amplitude for two particles. I will refer to the massive particles as $\chi$ and antiparticles as $\bar{\chi}$. From the second-order term of \eqref{dyson}, we have
\begin{align}
\begin{split}
\label{S2}
S^{(2)}_{\mb{p}_1 \mb{p}_2 \to \mb{p}'_1 \mb{p}'_2} & = - \int_{-\infty}^{\infty} dt_1 dt_2 d^3\mb{k}_1 d^3\mb{k}_2 V_{\mb{p}'_1 \mb{p}'_2,\mb{k}_1 \mb{k}_2} \delta^3(\mb{p}_1' + \mb{p}_2' - \mb{k}_1 -\mb{k}_2) V_{\mb{k}_1 \mb{k}_2,\mb{p}_1 \mb{p}_2} \\
& \times \delta^3(\mb{k}_1 + \mb{k}_2 - \mb{p}_1 - \mb{p}_2) \Theta(t_1-t_2) e^{-i (E_{\mb{p}'_1} + E_{\mb{p}'_2}) t_1} e^{i (E_{\mb{k}_1} + E_{\mb{k}_2}) (t_1-t_2)} e^{-i (E_{\mb{p}_1} + E_{\mb{p}_2}) t_2} \\
& + (t_1 \leftrightarrow t_2). 
\end{split}
\end{align}
To get the second line, we inserted a complete set of states $\ket{\mb{k}_1 \mb{k}_2}$. Now we insert an additional step function by noting that $\Theta(t) = \Theta^2(t)$, and use the same Fourier transform $\Theta(t) e^{-i E t} = (2\pi i)^{-1} \int d\omega e^{-i \omega t}/(\omega - E + i \epsilon)$ on each of the internal-line phase factors. Performing the time integrals, we obtain
\begin{align}
\begin{split}
\label{S2non-rel}
S^{(2)}_{\mb{p}_1 \mb{p}_2 \to \mb{p}'_1 \mb{p}'_2} & = \int d\omega_1 d^3\mb{k}_1 d\omega_2 d^3\mb{k}_2 V_{\mb{p}'_1 \mb{p}'_2,\mb{k}_1 \mb{k}_2} \delta(\omega_1 + \omega_2 - E_{\mb{p}'_1} - E_{\mb{p}'_2}) \delta^3(\mb{p}_1' + \mb{p}_2' - \mb{k}_1 -\mb{k}_2) \\
& \times \frac{1}{\omega_1 - E_{\mb{k}_1} + i\epsilon} \frac{1}{\omega_2 - E_{\mb{k}_2} + i\epsilon}  \\
& \times V_{\mb{k}_1 \mb{k}_2,\mb{p}_1 \mb{p}_2}\delta(\omega_1 + \omega_2 - E_{\mb{p}_1} - E_{\mb{p}_2}) \delta^3(\mb{k}_1 + \mb{k}_2 - \mb{p}_1 - \mb{p}_2)  \\ 
& + {\mb{p}_1 \mb{p}_2 \leftrightarrow \mb{p}'_1 \mb{p}'_2}.
\end{split}
\end{align}
One can interpret this as the amplitude for the initial state to transition to an intermediate state of two particles with momenta $\mb{k}_1, \mb{k}_2$ while conserving total energy and momentum, followed by free propagation of the two intermediate particles, followed by the transition to the final state $\ket{\mb{p}'_1 \mb{p}'_2}$. See the first panel of Fig. \ref{figure-chichibar}. The term labeled ${\mb{p}_1 \mb{p}_2 \leftrightarrow \mb{p}'_1 \mb{p}'_2}$, which comes from the $V(t_2) V(t_1)$ part of \eqref{S2}, is the same up to complex-conjugating the $V$ matrix elements.

Following the same logic as in previous sections, we promote this to a Lorentz-invariant amplitude:
\begin{align}
\begin{split}
\label{M2}
M_{\mb{p}_1 \mb{p}_2 \to \mb{p}'_1 \mb{p}'_2} & = \tilde{\alpha}^4 \int d^4k_1 d^4k_2 \delta^4(k_1+k_2 - p_1 - p_2) \\
& \times \frac{1}{k_1^2 + m^2 - i \epsilon} \frac{1}{k_2^2 + m^2 - i \epsilon} \frac{1}{(k_1-p_1)^2 + \mu^2 - i \epsilon} \frac{1}{(p'_1 - k_1)^2 + \mu^2 - i \epsilon}.
\end{split}
\end{align}
We could have done one of the integrals with the remaining delta-function to leave a single unconstrained loop integral, but it will become clear shortly why we write things this way. In the language of field theory, this is a box diagram with four scalars running in the loop. See Fig. \ref{figure-chichibar}. This brings up an important point: this expression is correct only in the very low-energy limit, i.e., the momentum integrals should have a ``UV'' cutoff $\Lambda \lesssim m$. Above $\Lambda$, we cannot know the detailed spectrum of the model (the intermediate states, non-relativistically included by inserting $1 = \sum \ket{\mb{k}_1 \mb{k}_2} \bra{\mb{k}_1 \mb{k}_2}$ above). However, we will only need the infrared part of this diagram in what follows, where the expression \eqref{M2} is reliable.

It is easy to check that \eqref{M2} satisfies the optical theorem. The imaginary part of the amplitude, which is of order $G_N^2$, is determined precisely by the square of the elastic tree-level $\chi \chi \to \chi \chi$ amplitude of order $G_N$. However, there will be a problem when we now extend these results to $\chi \bar{\chi}$ scattering. By crossing symmetry, the amplitude for $\chi \bar{\chi} \to \chi \bar{\chi}$ should be given by the $\chi \chi \to \chi \chi$ amplitude, but with the roles of some outgoing momenta switched with incoming momenta (see Fig. \ref{figure-chichibar}). At lowest order in perturbation theory, we simply get \eqref{M1rel}, but with the momentum transfer $t = -(p'_1 - p_1)^2$ replaced by the total incoming energy-momentum $s = -(p_1+p_2)^2$:
\be
M^{\chi \bar{\chi} \to \chi \bar{\chi}}_{\mb{p}_1 \mb{p}_2 \to \mb{p}'_1 \mb{p}'_2} = \frac{4\pi \tilde{\alpha}^2}{-s + \mu^2}.
\ee
For physical momenta $p_1^2 = p_2^2 = p_1'^2 = p_2'^2 = -m^2$, we have $s \geq 4m^2$. Thus the pole in the denominator cannot be reached, so the amplitude is purely real. Much like the $t$-channel $\chi\chi$ amplitude, the optical theorem is therefore trivial at this order since ${\rm Im}~M = 0 + \Oi(\tilde{\alpha}^4)$.

\begin{figure}[t]

\centering

\begin{tikzpicture}[scale=0.7, every node/.style={transform shape}]

\node [scale=1.2] at (-3.1,0) {Im};
\node [yscale=8.5,xscale=3] at (-2.5,0) {(};
\node [yscale=8.5,xscale=3] at (2.5,0) {)};

\draw [->-=.5] (-1.5,-2) -- (-1,-1);
\draw  [-<-=.5] (1.5,-2) -- (1,-1);
\draw [->-=.5] (-1,-1) -- (1,-1);
\draw [dashed] (-1,-1) -- (-1,1);
\draw [dashed] (1,-1) -- (1,1);
\draw [-<-=.5] (-1,1) -- (1,1);
\draw [-<-=.5] (1,1) -- (1.5,2);
\draw [->-=.5] (-1,1) -- (-1.5,2);

\node at (-1.5,2.3) {$\mb{p}'_1 = \mb{p}_1$};
\node at (1.5,2.3) {$\mb{p}'_2 = \mb{p}_2$};
\node at (-1.5,-2.3) {$\mb{p}_1$};
\node at (1.5,-2.3) {$\mb{p}_2$};

\node[scale=1.5] at (3.6,0) {$=$};

\node [yscale=3,xscale=2] at (4.5,0) {$\int$};
\node[scale=1.2] at (5.7,0) {$d^3\mb{k}_1 d^3\mb{k}_2$};

\begin{scope}[xshift=280]

\node [yscale=10] at (-2.6,0) {$|$};
\node [yscale=10] at (2.6,0) {$|$};
\node [scale=1.5] at (2.9,2.2) {$2$};

\draw [->-=.5] (-2,-1) -- (-1,0);
\draw [->-=.5] (-1,0) -- (1,0) ;
\draw [->-=.5] (1,0) -- (2,-1);
\draw [dashed] (1,0) -- (2,1);
\draw [dashed] (-2,1) -- (-1,0);

\node at (-2,1.3) {$\mb{k}_1$};
\node at (2,1.3) {$\mb{k}_2$};
\node at (-2,-1.3) {$\mb{p}_1$};
\node at (2,-1.3) {$\mb{p}_2$};

\end{scope}

\end{tikzpicture}

\caption{Forward $\chi \bar{\chi}$ scattering at second order in the gravitational interaction. The imaginary part of the forward amplitude has a branch cut singularity, which precisely equals the total cross section for the process $\chi \bar{\chi} \to {\rm gravitons}$.}
\label{figure-chichibaroptical}
\end{figure}

However, at the next order in perturbation theory $\Oi(\tilde{\alpha}^4)$, we will have unitarity problems. In particular, the amplitude will contain a term given by \eqref{M2} but with external momenta crossed:
\begin{align}
\begin{split}
\label{M2crossed}
M^{\chi \bar{\chi} \to \chi \bar{\chi}}_{\mb{p}_1 \mb{p}_2 \to \mb{p}'_1 \mb{p}'_2} & = \Oi(\tilde{\alpha}^2) +  \tilde{\alpha}^4 \int d^4k_1 d^4k_2 \delta^4(k_1+k_2 - p_1 - p_2) \\
& \times \frac{1}{k_1^2 + \mu^2 - i \epsilon} \frac{1}{k_2^2 + \mu^2 - i \epsilon} \frac{1}{(k_1-p_1)^2 + m^2 - i \epsilon} \frac{1}{(p'_1 - k_1)^2 + \mu^2 - i \epsilon}.
\end{split}
\end{align}
This looks almost identical to the part of the $\chi \chi \to \chi \chi$ amplitude given in \eqref{M2}, except that the roles of the particle mass $m^2$ and Yukawa scattering parameter $\mu^2$ are switched. To check unitarity, we can compute the imaginary part of the forward limit $\mb{p}_1' \to \mb{p}_1, \mb{p}_2' \to \mb{p}_2$ of this amplitude. A classic result of Cutkosky \cite{cutkosky1960singularities} says that the poles at $k_i^2 = -\mu^2$ combine to form an overall branch cut in the complex $s$-plane. The imaginary part of the diagram is computed as the complex discontinuity across this cut, and is given by replacing the two factors $1/(x + \mu^2 - i\epsilon) \to i \pi \delta(x + \mu^2)$. That is,
\be
\label{imM-chichibar}
{\rm Im}~M^{\chi\bar{\chi} \to \chi \bar{\chi}}_f(s) = - \pi^2 \tilde{\alpha}^4 \int \frac{d^3\mb{k}_1}{2 E_{\mb{k}_1}} \frac{d^3\mb{k}_2}{2 E_{\mb{k}_2}} \left| \frac{1}{(p_1 - k_1)^2 + m^2 - i \epsilon} \right|^2 \delta^4(k_1+k_2-p_1-p_2).
\ee
To satisfy the optical theorem, this would have to be given by something of the form of the right-hand side of \eqref{optical}. But much like the example of section \ref{section-main}, if the outgoing states are made up of $\chi$ and $\bar{\chi}$ particles alone, there is simply no set of diagrams that can satisfy this. In particular, the total elastic $\chi \bar{\chi}$ cross-section is given by
\be
\label{sigma-chichibar}
\sigma_{\chi \bar{\chi} \to \chi \bar{\chi}} = -\pi^2 \tilde{\alpha}^4  \int \frac{d^3\mb{k}_1}{2 E_{\mb{k}_1}} \frac{d^3\mb{k}_2}{2 E_{\mb{k}_2}} \left| \frac{1}{(p_1 - k_1)^2 + \mu^2 - i \epsilon} \right|^2 \delta^4(k_1+k_2-p_1-p_2). 
\ee
While these look similar, the problem is the mass in the denominators are different. Thus \eqref{optical} is not satisfied and we have a unitarity violation.

The solution is the same as before. We recognize that on the right-hand side of
\be
\label{opticalfinal2}
{\rm Im}~M_f = \sum_{X} | \braket{ X | S | \psi_{\rm in} } |^2
\ee
we need to add some new states $\ket{X}$. In this case, what we need to add is the process $\chi \bar{\chi} \to {\rm ``2~gravitons''}$. See Fig. \ref{figure-chichibaroptical}. Inserting the amplitude $M_{\chi \bar{\chi} \to gg}$ on the right hand side of \eqref{opticalfinal2} makes the equation precisely correct. Unlike the tree-level example, the sum is not over a single isolated point of phase space but rather a full integral over all possible kinematically allowed annihilations to two gravitons.

\end{appendix}

\bibliographystyle{utphys-dan} 
\bibliography{newton-implies-graviton}

\end{document}